\documentclass[%
 reprint,
 amsmath,amssymb,
 aps,
prd,
showkeys]{revtex4-2}
\usepackage{fix-cm}
\usepackage{amsfonts}
\usepackage{mathrsfs}
\usepackage{amsmath}
\usepackage{amssymb}
\usepackage{framed}

\usepackage{bm}

\usepackage[normalem]{ulem}
\usepackage{extarrows}
\usepackage{slashed}
\usepackage{isodateo}
\usepackage{graphicx}
\usepackage[dvipsnames]{xcolor}
\usepackage{indentfirst,comment}
\usepackage{booktabs}

\usepackage{bbm}
\usepackage{tikz-cd}
\usetikzlibrary{patterns}
\usepackage{tikz}
\usetikzlibrary{arrows,shapes}
\usetikzlibrary{trees}
\usetikzlibrary{matrix,arrows} 				
\usetikzlibrary{positioning}				
\usetikzlibrary{calc,through}				
\usetikzlibrary{decorations.pathreplacing}  
\usepackage{pgffor}							

\usetikzlibrary{decorations.pathmorphing}	
\usetikzlibrary{decorations.markings}
\tikzset{
	photon/.style={decorate, decoration={snake}, draw=red},
	electron/.style={draw=blue, postaction={decorate},
		decoration={markings,mark=at position .55 with {\arrow[draw=blue]{>}}}},
	gluon/.style={decorate, draw=blue,
		decoration={coil,amplitude=4pt, segment length=4pt}} ,
	vector/.style={decorate, decoration={snake}, draw},
	provector/.style={decorate, decoration={snake,amplitude=2.5pt}, draw},
	antivector/.style={decorate, decoration={snake,amplitude=-2.5pt}, draw},
	fermion/.style={draw=black, postaction={decorate},
		decoration={markings,mark=at position .55 with {\arrow[draw=black]{>}}}},
	fermionbar/.style={draw=black, postaction={decorate},
		decoration={markings,mark=at position .55 with {\arrow[draw=black]{<}}}},
	fermionnoarrow/.style={draw=black},
	scalar/.style={draw=black, postaction={decorate},
		decoration={markings,mark=at position .55 with {\arrow[draw=black]{>}}}},
	scalarbar/.style={dashed,draw=black, postaction={decorate},
		decoration={markings,mark=at position .55 with {\arrow[draw=black]{<}}}},
	scalarnoarrow/.style={dashed,draw=black},
	electron/.style={draw=black, postaction={decorate},
		decoration={markings,mark=at position .55 with {\arrow[draw=black]{>}}}},
	bigvector/.style={decorate, decoration={snake,amplitude=4pt}, draw},
	background/.style={dashed,draw=black, postaction={decorate},
		decoration={markings,mark=at position 1 with {\arrow[draw=black]{<>}}}},
}

\tikzstyle{block} = [draw, rectangle, 
minimum height=3em, minimum width=6em]

\definecolor{lightgreen}{cmyk}{0.2, 0, 0.2, 0.2}
\definecolor{lightgray}{cmyk}{0.1,0.2,0,0.1}
\definecolor{lightgray2}{cmyk}{0.1,0.1,0,0.1}
\definecolor{greyish2}{rgb}{.96,.96,.96}
\definecolor{bluecyan}{RGB}{0, 100, 200}
\definecolor{blue3}{RGB}{31,119,180}
\definecolor{red3}{RGB}{214,39,40}
\definecolor{orange3}{RGB}{255,127,14}
\definecolor{green3}{RGB}{44,160,44}
\definecolor{red2}{RGB}{255,0,0}
\definecolor{green2}{RGB}{0,170,0}
\definecolor{blue2}{RGB}{0,128,255}
\definecolor{magenta2}{RGB}{191,64,191}
\definecolor{purple2}{RGB}{112,48,160}
\definecolor{orange2}{RGB}{255,192,0}
\definecolor{blue2}{RGB}{117,223,230}
\definecolor{red4}{RGB}{186,60,71}

\newcommand{\n}{\nonumber}

\graphicspath{{fig/}}

\def\bge{\begin{equation}}
\def\ede{\end{equation}}
\def\bga{\begin{aligned}}
\def\eda{\end{aligned}}
\def\ie{\begin{equation}\begin{aligned}}
\def\fe{\end{aligned}\end{equation}}
\def\bgb{\begin{bmatrix}}
\def\edb{\end{bmatrix}}
\def\bgp{\begin{pmatrix}}
\def\edp{\end{pmatrix}}
\def\bgm{\begin{matrix}}
\def\edm{\end{matrix}}
\def\bgs{\begin{subequations}}
\def\eds{\end{subequations}}
\def\di{{\mathrm{d}}}

\def\mb{\mathbf}
\def\pd{\partial}

\def\ii{\mathrm{i}}

\def\ep{\epsilon}

\newcommand{\mG}{\mathcal{G}}
\newcommand{\mK}{\mathcal{K}}
\def\to{\rightarrow}

\def\2F1{{}_2\mathrm{F}_1}
\def\3F2{{}_3\mathrm{F}_2}

\usepackage{mdframed}

\newmdenv[skipabove=0pt,%
skipbelow=5pt,%
leftmargin=0pt,%
rightmargin=0pt,%
innertopmargin=-5pt,%
innerbottommargin=7pt,%
innerleftmargin=2pt,%
innerrightmargin=2pt,%
splittopskip=0pt,%
splitbottomskip=0pt,%
linewidth=0pt,%
nobreak=true]%
{keyeqn2}

\newmdenv[backgroundcolor=gray!15,%
skipabove=0pt,%
skipbelow=5pt,%
leftmargin=0pt,%
rightmargin=0pt,%
innertopmargin=-5pt,%
innerbottommargin=7pt,%
innerleftmargin=2pt,%
innerrightmargin=2pt,%
splittopskip=0pt,%
splitbottomskip=0pt,%
linewidth=0pt,%
nobreak=true]%
{keyeqn}
          
\newcommand{\fnemail}[1]{\footnote{Email: \href{mailto:#1}{\nolinkurl{#1}}}}

\usepackage{color}  
\definecolor{ceil}{rgb}{0.57, 0.63,0.81}

\begin{document}

\title{Loop integrals in de Sitter spacetime: The parity-split IBP system and $\mathrm{d}\log$-form differential equations}

\author{Jiaqi Chen$^{1,2}$}
\email{jiaqichen@cup.edu.cn}

\author{Bo Feng$^{3,4}$}
\email{fengbo@scnu.edu.cn}

\author{Zhehan Qin$^{5}$}
\email{qzh21@mails.tsinghua.edu.cn}

\author{Yi-Xiao Tao$^{6}$}
\email{taoyx21@mails.tsinghua.edu.cn}

\address{$^{1}$Beijing Key Laboratory of Optical Detection Technology for Oil and Gas, China University of Petroleum-Beijing, Beijing 102249, China\\
$^{2}$Basic Research Center for Energy Interdisciplinary, College of Science, China University of Petroleum-Beijing, Beijing 102249, China\\
$^{3}$State Key Laboratory of Nuclear Physics and Technology, Institute of Quantum Matter, South China Normal University, Guangzhou 510006, China\\
$^{4}$Guangdong Basic Research Center of Excellence for Structure and Fundamental Interactions of Matter, Guangdong Provincial Key Laboratory of Nuclear Science, Guangzhou 510006, China\\
$^{5}$Department of Physics, Tsinghua University, Beijing 100084, China\\
$^{6}$Department of Mathematical Sciences, Tsinghua University, Beijing 100084, China}

\begin{abstract}
We develop integration-by-parts (IBP) reduction and differential equations for massive loop integrals of cosmological correlators in de Sitter (dS) spacetime, demonstrating the feasibility of this approach. We identify a structural property of the dS IBP system: for an $n$-propagator family, it splits into $2^n$ closed subsystems classified by the parity of the propagator indices. We further formulate a Baikov representation for loop integrals in dS space and derive the corresponding dimensional recurrence relations. In flat spacetime, intersection theory shows that $\mathrm{d}\log$-form master integrands lead to $\mathrm{d}\log$-form differential equations. Motivated by fibration intersection theory, we conjecture that this construction extends to dS integrands involving Hankel functions. We verify this conjecture in the one-loop bubble family and determine the associated alphabet.
\end{abstract}

\keywords{Cosmology, Scattering amplitudes, Quantum field theory in curved spacetime, Feynman integrals}

\maketitle

\section{Introduction}

De Sitter (dS) spacetime is of central phenomenological importance in inflationary cosmology \cite{Achucarro:2022qrl} and at the same time provides the simplest curved background for developing systematic perturbative methods. For massless or conformally coupled fields, the integrands simplify considerably and often retain structures closely analogous to the polynomial-type integrands familiar from flat space \cite{Arkani-Hamed:2024jbp,Arkani-Hamed:2023kig,Arkani-Hamed:2023bsv,Benincasa:2022gtd,Benincasa:2024leu,Benincasa:2024lxe,Fan:2024iek,Benincasa:2024ptf,Baumann:2024mvm,He:2024olr,Chowdhury:2025ohm,De:2023xue,De:2025bmf,Glew:2025otn,Glew:2025ypb,Pimentel:2026kqc,Hang:2024xas,Jain:2025maa,Farren:2026hao}. This has led both to amplitude-inspired developments \cite{Arkani-Hamed:2017fdk,Arkani-Hamed:2018bjr,Lee:2022fgr,Gomez:2021qfd,Gomez:2021ujt,Armstrong:2022mfr,Tao:2022nqc,Chen:2023bji,Chen:2023xlt} and to direct extensions of flat-space techniques such as integration-by-parts (IBP) reduction \cite{Chetyrkin:1981qh} and differential equations \cite{Kotikov:1990kg,Kotikov:1991pm,Gehrmann:1999as,Bern:1993kr}. For correlators with massive intermediate states, several powerful complementary analytic approaches have also been proposed, including the cosmological bootstrap \cite{Arkani-Hamed:2018kmz,Baumann:2019oyu,Baumann:2020dch,Pajer:2020wnj,Pajer:2020wxk,Cabass:2021fnw,Baumann:2022jpr,Pimentel:2022fsc,Jazayeri:2022kjy,Qin:2022fbv,Wang:2022eop,Qin:2023ejc,Aoki:2024uyi,Liu:2024str,
Qin:2025xct,Xianyu:2025lbk}, Mellin techniques \cite{Sleight:2019mgd,Sleight:2019hfp,Sleight:2020obc,Sleight:2021plv,Premkumar:2021mlz,Qin:2022lva,Qin:2022fbv,Xianyu:2023ytd,Qin:2024gtr}, spectral decompositions \cite{Marolf:2010zp,Xianyu:2022jwk,Loparco:2023rug,Werth:2024mjg,
Qin:2025xct,Zhang:2025nzd,Altshuler:2025qmk} and dispersive methods \cite{Liu:2024xyi,Das:2026vfv}. See also \cite{Wang:2021qez,Werth:2023pfl,Pinol:2023oux,Werth:2024aui,Chen:2024glu} for some recent numerical studies.

Nevertheless, explicit computations of such correlators, which are central to cosmological collider physics \cite{Chen:2009we,Chen:2009zp,Baumann:2011nk,Noumi:2012vr,Arkani-Hamed:2015bza}, remain largely limited to tree level and to bubble topologies at one loop. Among currently available analytic tools, only spectral decompositions \cite{Xianyu:2022jwk,Zhang:2025nzd,Cespedes:2025ple} and the partial Mellin-Barnes representation \cite{Qin:2024gtr} have been successfully applied to one-loop bubble diagrams, while extending them to other important one-loop topologies, such as triangles and boxes, appears less straightforward. Existing results beyond bubbles include triangles of conformally coupled or massless scalars \cite{Pimentel:2026kqc}, a special case in which the triangle arises from a free theory via field redefinitions \cite{Wang:2025qfh}, and partial results for triangle and box diagrams obtained from partial Mellin-Barnes representations \cite{Qin:2023bjk}. By contrast, IBP reduction is organized at the level of integral families and is therefore naturally suited for systematic generalization.

From the technical viewpoint, the massive case is substantially more challenging because the relevant integrands involve Hankel functions and are no longer of polynomial type. Nevertheless, Ref.~\cite{Chen:2023iix} pointed out that IBP reduction and the method of differential equations can still be extended to massive dS correlators. At tree level, and for the time-integral part at loop level, this leads to differential equations in \(\di\log\)-form \cite{Chen:2024glu,Baumann:2026atn,Gasparotto:2024bku} and new analytic solutions in terms of hypergeometric functions \cite{Chen:2024glu}. 

In this letter, we take a first step toward the complete loop-level massive case, including both the time and loop-momentum integrals.
We formulate the IBP reduction for this family and show that the reduction system decomposes into independent sectors labeled by the odd/even parity of the indices associated with each propagator. More generally, for an \(n\)-propagator family, this implies a decomposition into \(2^n\) IBP-closed subsystems, greatly simplifying the reduction of dS loop integrals.

We also investigate the analytic structure of the associated differential equations. In flat spacetime, canonical differential equations \cite{Henn:2013pwa} in \(\di\log\)-form play a central role in the analytic computation of Feynman integrals. It has long been observed that \(\di\log\)-form (or leading-singularity) master integrands \cite{Arkani-Hamed:2012zlh, Bern:2014kca, Chicherin:2018old, Chen:2020uyk, Chen:2022lzr,Henn:2020lye, Dlapa:2021qsl} lead to \(\di\log\)-form differential equations, and intersection theory \cite{Mastrolia:2018uzb,Frellesvig:2019uqt,Mizera:2019ose,Chestnov:2022xsy,Weinzierl:2020xyy,Chen:2020uyk,Chen:2022lzr,Jiang:2023oyq,Crisanti:2024onv,Frellesvig:2020qot,Mizera:2019vvs,Chestnov:2022alh,Caron-Huot:2021xqj,Caron-Huot:2021iev,Giroux:2022wav,Duhr:2024rxe,Fontana:2023amt,Brunello:2023rpq,Brunello:2024tqf,Huang:2026xnq} provided a natural framework for understanding this mechanism \cite{Chen:2023kgw,Chen:2024ovh}.
Motivated by \cite{Chen:2023kgw,Chen:2024ovh} and the fibration approach to multivariate intersection numbers \cite{Frellesvig:2019uqt}, we propose a generalized perspective adapted to the massive dS case, in which the master integrals of the \(\tau\)-integrated kernel are treated as a twist \(\mathbf U\), while the remaining part of the integrand is encoded in a differential form \(\Phi\). This leads to a practical criterion for obtaining \(\di\log\)-form differential equations through an a priori construction of the integrand.

Using a dS version of the Baikov representation \cite{Baikov:1996iu}, we implement this construction for the one-loop bubble family, verify that the homogeneous top-sector block of the differential-equation
system is in \(\di\log\)-form, and determine the corresponding alphabet explicitly. Taken together, these results provide evidence that loop-level massive cosmological correlators in dS space admit a systematic treatment that parallels the flat-space case despite the non-polynomial structure of the integrands.

\section{Setup}
\label{sec:definitions}

\begin{figure}[ht]
    \centering
\begin{tikzpicture}[baseline={([yshift=-.5ex]current bounding box.center)}, line width=1. pt, scale=3.]
    \draw[blue3, thick] (0.25,0) ellipse (0.25 and 0.20);
    \draw[black] (0, 0) -- (-0.285, 0.305);
    \draw[black] (0, 0) -- (-0.285, -0.305);
    \node[draw, rectangle, minimum width=6.0pt, minimum height=6.0pt, inner sep=0pt, anchor=south east] at (-0.28,0.3) {};
    \node[draw, rectangle, minimum width=6.0pt, minimum height=6.0pt, inner sep=0pt, anchor=north east] at (-0.28,-0.3) {};
    \draw[black] (0.5, 0) -- (0.755, 0.305);
    \draw[black] (0.5, 0) -- (0.755, -0.305);
    \draw[draw=lightgray2, fill=lightgray2] (0, 0) circle (.035cm);
    \draw[draw=lightgray2, fill=lightgray2] (0.5, 0) circle (.035cm);
    \node[draw, rectangle, minimum width=6.0pt, minimum height=6.0pt, inner sep=0pt, anchor=south west] at (0.75,0.3) {};
    \node[draw, rectangle, minimum width=6.0pt, minimum height=6.0pt, inner sep=0pt, anchor=north west] at (0.75,-0.3) {};
    \node at (-0.08, 0.23) {$k_1$};
    \node at (-0.08, -0.23) {$k_2$};
    \node at (0.58, 0.23) {$k_3$};
    \node at (0.58, -0.23) {$k_4$};
    \node at (0.25,0.36){\textcolor{blue3}{$\mb q$}};
    \node at (0.25,-0.36){\textcolor{blue3}{$\mb q+\mb k_s$}};
    \draw[<-,blue3] (0.06,0.27)--(0.43,0.27);
    \draw[->,blue3] (0.06,-0.27)--(0.43,-0.27);
\end{tikzpicture}
    \caption{SK diagrams for four-point correlators with bubble topology. The external lines represent conformally coupled scalars, while the blue loop lines represent massive scalars. The \(\Box\) vertices lie on the future boundary, and the \textcolor{lightgray2}{$\bullet$} vertices denote bulk insertions on the two Schwinger-Keldysh contours, with \(+\) and \(-\) corresponding to the time-ordered and anti-time-ordered branches, respectively. Contributions from all SK branches are summed.\label{fig_bubble}}
\end{figure}
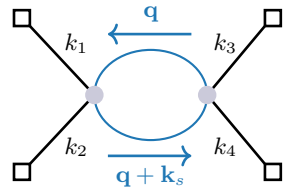

We work in the Poincar\'e patch of de Sitter space, with conformal time \(\tau\in(-\infty,0)\), where the metric takes the form
\begin{equation}
\di s^2=a^2(\tau)\bigl(-\di\tau^2+\di\mb x^2\bigr)\,,\qquad
a(\tau)=-\frac{1}{H\tau}\,.
\end{equation}
For convenience, we set \(H=1\) throughout. Although most of our discussion applies to general spatial dimension \(d\), the cosmologically relevant case is \(d=3\); dimensional regularization is implemented by analytic continuation to \(d=3-2\ep\).
Following Ref.~\cite{Chen:2023iix}, we introduce the rescaled Hankel building blocks \footnote{For scalar particles, the Hankel index \(\nu\) is purely imaginary for the principal series and real for the complementary series. In both cases, \(\mathrm H_{\nu^*}^{(2)}(z)\propto \mathrm H_\nu^{(2)}(z)\). As a result, although \(h_\nu^{(2)}\) was defined in Ref.~\cite{Chen:2023iix} using \(\mathrm H_{\nu^*}^{(2)}\), the present definition is valid for arbitrary \(\nu\) and differs from that of Ref.~\cite{Chen:2023iix} only by an overall normalization.}
\begin{align}
h_\nu^{(\alpha)}(z) \equiv z^{-\nu}\mathrm H_\nu^{(\alpha)}(z)\,,\qquad \alpha=1,2\,,\quad
\nu\in\mathbb C\,,
\label{eq:h}
\end{align}
which satisfy the rescaled Bessel equation
\begin{equation}
\partial_z^2 h_\nu^{(\alpha)}(z)+\frac{2\nu+1}{z}\partial_z h_\nu^{(\alpha)}(z)+h_\nu^{(\alpha)}(z)=0\,.
\label{eq:hdef}
\end{equation}
When the label \(\nu\) is clear from the context, we suppress it and denote derivatives by
\begin{align}
\partial_z^n h_\nu^{(\alpha)}(z)\equiv h_n^{(\alpha)}(z)\,,
\end{align}
Once $h_2$ appears, we will reduce it to $h_0$ and $h_1$ immediately by \eqref{eq:hdef}.
We use the standard Schwinger-Keldysh (SK) formalism \cite{Schwinger:1960qe,Feynman:1963fq,Keldysh:1964ud,Weinberg:2005vy} to compute equal-time correlators, following the diagrammatic rules of Ref.~\cite{Chen:2017ryl}. As a first example, we consider the 1-loop bubble diagram with four external legs of the conformally coupled scalar \(\phi_{\text{c}}\) (while the minimally coupled case shares the same function family) mediated by a loop of massive scalars \(\sigma\), as shown in Fig.~\ref{fig_bubble}.
\bigskip

The main technical difficulty lies in the bulk-to-bulk propagators,
\begin{align}
G_{\pm\pm}(k;\tau_1,\tau_2)=&~\theta_{12}G_{\mp\pm}(k;\tau_1,\tau_2) +\theta_{21}G_{\pm\mp}(k;\tau_1,\tau_2)\,, \notag\\
G_{-+}(k;\tau_1,\tau_2)\equiv&~u(k,\tau_1)u^*(k,\tau_2)\,, \notag\\
G_{+-}(k;\tau_1,\tau_2)\equiv&~u^*(k,\tau_1)u(k,\tau_2)\,, 
\label{eq:proptype}
\end{align}
where the mode function and its complex conjugate are
\begin{align}
u(k,\tau)&=\frac{\sqrt\pi}{2}e^{\ii\pi \nu/2}(-\tau)^{d/2}\mathrm H_{\nu}^{(1)}(-k\tau)\,, \notag\\
u^*(k,\tau)&=\frac{\sqrt\pi}{2}e^{-\ii\pi \nu^*/2}(-\tau)^{d/2}\mathrm H_{\nu^*}^{(2)}(-k\tau)\,.
\end{align}
Here \(\nu\equiv\sqrt{d^2/4-m^2}\) is the mass parameter. For notational simplicity, we focus on the complementary-series case where \(\nu\) is real, while our method applies equally well to the principal-series case where \(\nu\) is purely imaginary. We also use the shorthand \(\theta_{ij}\equiv\theta(\tau_i-\tau_j)\). By contrast, the bulk-to-boundary propagators of \(\phi_{\text{c}}\) take the simple form
\begin{align}
K_\pm(k;\tau)= \frac{\tau}{2k}e^{\pm \ii k\tau}\,.
\end{align}

We denote by \(\mathcal T_{\mathsf a\mathsf b}\), with \(\mathsf a,\mathsf b\in\{+,-\}\), the contribution from the SK branch in which the vertices at \(\tau_1\) and \(\tau_2\) lie on the \(\mathsf a\) and \(\mathsf b\) contours, respectively. Of the four branches, \(\mathcal T_{++}\) and \(\mathcal T_{--}\) are more involved because \(G_{++}\) and \(G_{--}\) each contain both \(\theta_{12}\) and \(\theta_{21}\) terms. Since \(\mathcal T_{++}=\mathcal T_{--}^*\), it suffices to consider \(\mathcal T_{--}\). For the simplest contact interaction \(a^{d+1}\phi_{\text{c}}^2\sigma^2\), one finds 
\begin{align}\label{eq_Feynmanrule}
\mathcal T_{--} \propto&~ \int_{-\infty}^0 \frac{\di \tau_1}{(-\tau_1)^{d+1}}\frac{\di \tau_2}{(-\tau_2)^{d+1}} \notag\\
&\times K_-(k_1;\tau_1)K_-(k_2;\tau_1)K_-(k_3;\tau_2)K_-(k_4;\tau_2) \notag\\
&\times \int \frac{\di^d\mb q}{(2\pi)^d}\,G_{--}(|\mb q|;\tau_1,\tau_2)G_{--}(|\mb q+\mb k_s|;\tau_1,\tau_2)\,.
\end{align}
Here \(k_i\) is the magnitude of \(\mb{k}_i\). This motivates the following integral family for the bubble topology, which accommodates vertices with arbitrary numbers of derivatives:
\begin{widetext}
\ie\label{eq_Idef2}
&I[\{\bm n\},\{a_1,a_2\},\{b_1,b_2\}]
=\int \frac{\di^d \mb q}{(2\pi)^d}
\frac{1}{|\mb q|^{b_1}|\mb k_s+\mb q|^{b_2}} I^{\tau}_{ \{\bm n, a_1, a_2\} } \,,\quad\quad I^{\tau}_{\{\bm n, a_1, a_2\}}  \equiv \int_{-\infty}^0 \di\tau_1\di\tau_2\,\hat{I}^{\tau}_{\{\bm n, a_1, a_2\}} \,, \n \\
& \hat{I}^{\tau}_{\{\bm n, a_1, a_2\}}=e^{-\ii P_1\tau_1 - \ii P_2 \tau_2}\,(-\tau_1)^{a_1}(-\tau_2)^{a_2} \left[ \theta_{21}h_{n_1}^{(1)}(-|\mb q|\tau_1)h_{n_2}^{(2)}(-|\mb q|\tau_2)
+\theta_{12}h_{n_1}^{(2)}(-|\mb q|\tau_1)h_{n_2}^{(1)}(-|\mb q|\tau_2)\right] \\
&~~~~~~~~~~~~~\times \left[\theta_{21}h_{n_3}^{(1)}(-|\mb q+\mb k_s|\tau_1)h_{n_4}^{(2)}(-|\mb q+\mb k_s|\tau_2)
+\theta_{12}h_{n_3}^{(2)}(-|\mb q+\mb k_s|\tau_1)h_{n_4}^{(1)}(-|\mb q+\mb k_s|\tau_2)\right]\,. 
\fe
\end{widetext}

Here \(P_1=k_1+k_2\) and \(P_2=k_3+k_4\) are the energies flowing into the two vertices. Equation~\eqref{eq:hdef} implies that any \(h_n^{(\alpha)}(z)\) with \(n\ge 2\) can be recursively reduced to a linear combination of \(h_{0}^{(\alpha)}(z)\) and \(h_{1}^{(\alpha)}(z)\). It is thus sufficient to restrict all indices \(n_i\) to \(\{0,1\}\). The contribution in \eqref{eq_Feynmanrule} then corresponds, up to an overall normalization, to
\begin{equation}\label{eq_TinI}
\mathcal T_{--} \propto I[\{0,0,0,0\},\{1+2\nu,1+2\nu\},\{-2\nu,-2\nu\}]\,.
\end{equation}

The family \eqref{eq_Idef2} is manifestly closed under IBP identities generated by total derivatives in the loop momentum \(\mb q\). Temporal IBP is more subtle, because derivatives acting on \(\theta_{21}\) and \(\theta_{12}\) generate contact terms proportional to \(\delta(\tau_1-\tau_2)\). As explained in Ref.~\cite{Chen:2023iix}, after using this delta function to localize one of the time integrals, the collapsed propagator is evaluated through the Wronskian relation for real $\nu$:
\begin{align}
h_{n_1}^{(1)}(z)\,h_{n_2}^{(2)}(z)-h_{n_1}^{(2)}(z)\,h_{n_2}^{(1)}(z)
= (n_1-n_2)\frac{4\ii}{\pi}z^{-2\nu-1}\,,
\label{eq_propcontract}
\end{align}
with \(n_{1,2}\in\{0,1\}\). It follows that, under \(\di\tau\)-IBP, the family \eqref{eq_Idef2} closes up to residual terms belonging to the tadpole-like family 
\begin{align}\label{eq_Rdef}
& ~~~~ R[\{n_1,n_2\},\{a\},\{b_1,b_2\}] \n\\
 &\equiv  \frac{4\ii}{\pi}\int_{-\infty}^0 \di\tau \int \frac{\di^d \mb q}{(2\pi)^d}  \frac{e^{-\ii P_0\tau} (-\tau)^{a}}{|\mb q|^{b_1}|\mb k_s+\mb q|^{b_2}}  \n\\
& \times \frac{1}{2}\sum_{i=0,1} h_{n_1}^{(1+i)}(-|\mb q|\tau)h_{n_2}^{(2-i)}(-|\mb q|\tau) 
\end{align}
where \(P_0\equiv P_1+P_2\). When \(b_2=0\), \(R\) reduces to the standard tadpole family.

For comparison, \(\mathcal T_{-+}\), in which the first bulk vertex lies on the \(+\) contour and the second on the \(-\) contour, is naturally encoded in the family
\ie\label{eq_Ipmdef}
& \widetilde I[\{\bm{n}\},\{a_1,a_2\},\{b_1,b_2\}] = \int_{-\infty}^0 \di\tau_1\di\tau_2 \int \frac{\di^d \mb q}{(2\pi)^d}\,\\
&\times  e^{-\ii P_1\tau_1+\ii P_2 \tau_2}\,
	\frac{(-\tau_1)^{a_1}(-\tau_2)^{a_2}}{|\mb q|^{b_1}|\mb k_s+\mb q|^{b_2}} h_{n_1}^{(1)}(-|\mb q|\tau_1)h_{n_2}^{(2)}(-|\mb q|\tau_2) \\
&\times h_{n_3}^{(1)}(-|\mb q+\mb k_s|\tau_1)h_{n_4}^{(2)}(-|\mb q+\mb k_s|\tau_2)\,.
\fe
Similar to \eqref{eq_TinI}, we have
\begin{equation}
\mathcal T_{-+} \propto \widetilde I[\{0,0,0,0\},\{1+2\nu,1+2\nu\},\{-2\nu,-2\nu\}]\,.
\end{equation}
Since \eqref{eq_Ipmdef} contains no step functions, it closes under both \(\di q\)-IBP and \(\di\tau\)-IBP, with no additional tadpole-like family. Its IBP relations follow from those of \eqref{eq_Idef2} upon replacing \(P_2\to -P_2\) and discarding all terms proportional to \(R\). We therefore restrict attention to the family \eqref{eq_Idef2}. The complete IBP relations are given in the Supplemental Material. The top-sector relations of \eqref{eq_Idef2} immediately reveal the following conserved \emph{parities}:
\begin{enumerate}
\item For both \(\di q\)-IBP and \(\di \tau\)-IBP, the parities of \(n_1+n_2+b_1\) and \(n_3+n_4+b_2\) are conserved. Each conserved parity is naturally associated with one propagator.
\item For \(\di q\)-IBP, the parities of \(n_1+n_3+a_1\) and \(n_2+n_4+a_2\) are also conserved \footnote{Since our reduction combines \(\di q\)-IBP and \(\di \tau\)-IBP, this second pair will not be used below.}.
\end{enumerate}
In particular, in the residual family \(R\) defined in \eqref{eq_Rdef}, the line carrying momentum \(\mb k_s+\mb q\) no longer contains any Hankel functions, so the corresponding conserved parity is simply the parity of \(b_2\).

More generally, the first conservation pattern extends straightforwardly to \(n\)-propagator families, yielding \(2^n\) IBP-closed subsystems and thus significantly simplifying the reduction of dS loop integrals. 
For simplicity, in the following part, we restrict attention to the even-even subsystem with $P_1=P_2$:
\begin{align}
n_1+n_2+b_1 \ \text{even}\,,\qquad n_3+n_4+b_2 \ \text{even}\,.
\label{eq:parity_even_sector}
\end{align}
We reduce this subsystem using the user-defined system module of Kira
\cite{Maierhofer:2017gsa,Klappert:2020nbg,Lange:2025fba}.
It contains 14 top-sector master integrals and 5 in the remaining-term
family. The five remaining-term masters are genuine members of the
closed even-even system: temporal-IBP contact terms allow derivatives
of top-sector masters to contain them. The construction below concerns
the homogeneous top-to-top block of this complete system.

\section{\(\di\log\) integrands of the top sector and the alphabet}
\label{sec:baikov}

In this section, we construct the \(\di\log\)-form integrands for the top sector and determine the alphabet of the associated differential system. Since the Baikov representation is particularly well suited to the construction of \(\di\log\)-forms in flat spacetime, we first adapt it to cosmological correlators and then explain how a viewpoint inspired by fibration intersection theory motivates our construction.

For the dS bubble diagram, we introduce the Baikov variables \cite{Baikov:1996iu}
\begin{align}
x_1=\sqrt{z_1}=|\mb q|\,,\qquad x_2=\sqrt{z_2}=|\mb q+\mb k_s|\,.
\end{align}
Using the same change of variables as in the flat-space Baikov representation, one finds the normalized measure
\ie
&\frac{2}{\pi^{d/2}}\,\di^d q
= C_d\,\mG^{\frac{d-3}{2}}\mK^{-\frac{d-2}{2}}\,\di z_1\,\di z_2\,,\ 
C_{d} = \frac{\pi^{-1/2}}{\Gamma[(d-1)/2]}\,, \\
&\mK=k_s^2\,, \ 
\mG=\frac{1}{4}\Big(2 z_1 k_s^2+2 z_2 k_s^2-k_s^4-z_1^2-z_2^2+2 z_1 z_2\Big)\,,
\fe
valid for general spatial dimension \(d\). For later use, note that \(\di z_i=2x_i\,\di x_i\), so that a factor \(x_i^{-b_i}\) in the \(z_i\) representation is shifted by one power when rewritten in terms of \(x_i\):
\begin{align}
&~I[\{\bm{n}\},\{a_1,a_2\},\{b_1,b_2\}]
\sim
\int \cdots \times \frac{\di z_1}{x_1^{b_1}}\frac{\di z_2}{x_2^{b_2}}\n\\
=&~
\int \cdots \times 4\,\frac{\di x_1}{x_1^{b_1-1}}\frac{\di x_2}{x_2^{b_2-1}}\,.
\end{align}
The Baikov representation also allows the dimensional recurrence relations \cite{Tarasov:1996br,Lee:2010wea} to be carried over from flat spacetime to dS space; the relevant relation is given in \eqref{shift}.

Since IBP shifts the indices \(a_i\) and \(b_i\) only by integers, it is convenient to absorb the common \(\nu\)-dependent offsets into the index notation and rewrite
\begin{align}
	&I[\{\bm{n}\},\{a_0+a_1',a_0+a_2'\},\{b_0+b_1',b_0+b_2'\}]\notag\\
	\to &I[\{\bm{n}\},\{a_1',a_2'\},\{b_1',b_2'\}]
\end{align}
where, as seen from \eqref{eq_TinI}, these shifts are
\begin{align}
	a_0 = 2\nu  \,, \quad b_0= -2\nu \,.
\end{align}

We now briefly recall the relevant aspect of intersection theory. In flat spacetime, a Baikov integrand can be written as \(u\,\phi\), with \(u=\prod_i P_i^{\beta_i}\) and $\phi$ a $\di\log$ form, for example \(u=\mG^{-\ep}\), so that the integrand is built entirely from polynomial factors. The associated twisted connection,
\begin{align}
\omega \equiv \frac{\di u}{u}=\sum_i \beta_i\,\di\log P_i\,,
\end{align}
is therefore naturally of \(\di\log\)-form. It was further shown in \cite{Chen:2023kgw,Chen:2024ovh} that \(\di\log\)-form master integrands \(\{\phi_i\}\) with \(\di\log\)-form \(\omega\) lead to \(\di\log\)-form differential equations.

Intersection theory is usually formulated for polynomial-type integrands, whereas the dS integrands relevant here involve Hankel functions. To address this, we reinterpret the intermediate step in the fibration approach to intersection theory \cite{Frellesvig:2019uqt} as suggesting a generalized intersection-theoretic structure acting on objects of the form \(\Phi\cdot\mathbf U\). Here \(\mathbf U\) denotes a vector of master integrals for a kernel family, while \(\Phi\) collects the remaining part of the integrand. Unlike in the standard setting, we do not require the integrand defining \(\mathbf U\) to be of polynomial type.

For dS loop integrals, the \(\tau\) integral can be incorporated into the kernel \(\mathbf U\). Unlike in flat spacetime, the twisted connection \(\Omega\), defined by
\begin{align}
\di \mathbf U=\Omega\cdot\mathbf U\,,
\end{align}
encodes the differential equations satisfied by the kernel integrals and is not automatically \(\di\log\) form. We conjecture that, if \(\mathbf U\) is chosen such that \(\Omega\) is in \(\di\log\) form and the remaining part of the integrands \(\{\Phi\}\) are also chosen to be in \(\di\log\) form, then the corresponding differential equations are automatically in \(\di\log\)-form. We implement this idea below.

We choose the kernel to be
\begin{align}\label{un}
	\mathbf{U}_{\bm n} = x_1^{-b_0}x_2^{-b_0}\,\mG^{-\ep}\mK^{\ep}\,I^{\tau}_{\bm n}\,,
\end{align}
where \(\ep\) is the dimensional regulator and \(d=3-2\ep\). The one-loop Baikov measure contributes the factor \(\mG^{-\ep}\mK^{\ep}\), and \(I^{\tau}_{\bm n}\), with all \(n_i\) taking values 0 or 1, labels the master integrals of the \(\tau\)-integral family, as studied in \cite{Chen:2023iix}. 

Ref.~\cite{Chen:2023iix} also shows that the differential equations for \(I^{\tau}_{\bm n}\) are in \(\di\log\)-form
\begin{align}
	\di I^{\tau}_{\bm m}=(\di\Omega_{\bm m\bm n}^{\tau})\,I^{\tau}_{\bm n} \,, \quad \di\Omega^{\tau}=\di\Omega_{x_1}+\di\Omega_{x_2}+\di\Omega_{ex}\,,  \label{eq:Omegatau}
\end{align}
where the one-forms in \(\di\Omega^{\tau}\) are built from \(\di\log x_1\), \(\di\log x_2\), and \(\di\log(P_i\pm x_1\pm x_2)\), and the three contributions are denoted by \(\di\Omega_{x_1}\), \(\di\Omega_{x_2}\), and \(\di\Omega_{ex}\), respectively. 
Accordingly, the twisted connection associated with \(\mathbf{U}\) is also of \(\di\log\) form,
\begin{align}
	\di \mathbf{U}=&~ \Omega \cdot \mathbf{U}\notag\\
 =&\Big(\di\Omega^{\tau}-b_0(\di\log x_1+\di\log x_2)\notag\\
 &~~~~~~~~~~~~~~~~~-\ep(\di\log \mG-\di\log \mK)\Big)\cdot \mathbf{U} \,.
\end{align}

For fixed \(z_2\), the two zeros of \(\mG\), viewed as a
polynomial in \(z_1\), are
\(z_1=(k_s\pm\sqrt{z_2})^2\), so we have the following useful building blocks:
\begin{align}\label{eq_buildingblock}
	\di\log z_1&=2\di\log x_1\, ,\notag\\
	\di\log \left(\frac{(k_s + \sqrt{z_2})^2 - z_1}{(k_s - \sqrt{z_2})^2 - z_1}\right) &= -\frac{2k_sx_1x_2\,\di x_1}{\mG}\notag\\
 &~~~-\frac{k_s^3-k_sx_1^2-k_sx_2^2}{\mG}\di x_2
\end{align}
and $(z_1\leftrightarrow z_2)$. The connection \(\di\Omega^\tau\) contains additional \(\di\log\)-type building blocks. However, some of them involve denominators of the form \(P_i \pm x_1 \pm x_2\), and using such factors directly would enlarge the integral family. To avoid this, we only use combinations that can be rewritten back in terms of the original family:
\begin{align}
	&\partial_{P_j}\Omega_{ex}\,\di x_i = \partial_{P_j}\Omega^{\tau}\,\di x_i 
 \, , \n\\
	&\partial_{x_j}\Omega_{ex}\,\di x_i = (\partial_{x_j}\Omega^{\tau}-\partial_{x_j}\Omega_{x_j})\,\di x_i \,,
\end{align}
When acting on \(I^\tau_{\bm n}\), the first relation amounts,
up to an overall constant, to a \(\tau_j\) insertion. The second
can be mapped back to the original family by an \(x_j\)-IBP,
while \(\partial_{x_j}\Omega_{x_j}\) contains only \(1/x_j\) and
therefore introduces no new denominator structure. Further details
are given in the Supplemental Material.

Pairing these \(x_1\)- and \(x_2\)-fiber one-forms, we obtain the
following five classes of logarithmic two-forms. Their representatives
in the original family are obtained from
\(\int\Phi_{\bm m\bm n}\mathbf U_{\bm n}\) with $\mathbf U$ given in \eqref{un}.
\begin{enumerate}
	\item[(a)] The simplest scalar pairing combines the Gram logarithmic one-form in \eqref{eq_buildingblock} with $\di\log x_2$, namely, $\Phi_{\bm m\bm n}\propto \delta_{\bm m\bm n}\,\frac{x_1x_2k_s\,\di x_1}{\mG}\frac{\di x_2}{x_2}$. This gives:
	\begin{align}
	&I[\{0, 0, 0, 1\}, \{0, 0\}, \{0, 1\}, d=1-2\ep]\,,\notag\\
	&I[\{1, 1, 0, 1\}, \{0, 0\}, \{0, 1\}, d=1-2\ep]\,.
 \end{align}
\item[(b)] The kernel pairing combines the complete Gram
logarithmic one-form in \eqref{eq_buildingblock} with the complete
$P_2$-differentiated kernel one-form displayed in the first line
of \eqref{dOmega}. Rewriting the resulting logarithmic two-form
in the original integral family gives \footnote{This requires additional $\di\tau$-IBP.}:
\begin{align}
&I[\{0,0,0,0\},\{0,1\},\{0,0\}]
 +\frac12\mathcal C_{\{0,1,0,1\}},
\notag\\
&I[\{0,0,1,1\},\{0,1\},\{0,0\}]
 -\frac12\mathcal C_{\{0,1,1,0\}},
\notag\\
&I[\{1,1,1,1\},\{0,1\},\{0,0\}]
 +\frac12\mathcal C_{\{1,0,1,0\}},\notag\\
\mathcal C_{\bm n}\equiv {}&
 k_s^2 I[\{\bm n\},\{0,1\},\{1,1\}]-I[\{\bm n\},\{0,1\},\{-1,1\}]\notag\\
&\qquad~-I[\{\bm n\},\{0,1\},\{1,-1\}].
\label{eq:class-b-integral}
\end{align} 
Here all integrals are evaluated at $d=1-2\ep$.
	\item[(c)] Replacing the Gram form in class (a) by $\di\log x_1$ gives the
second scalar pairing: $\Phi_{\bm m\bm n} \propto \delta_{\bm m\bm n}\,\frac{\di x_1\,\di x_2}{x_1 x_2}$, and thus
\begin{align}
	& k_sI[\{0, 0, 0, 0\}, \{0, 0\}, \{2, 2\}, d=3-2\ep]\,, \notag\\
	&k_sI[\{0, 0, 1, 1\}, \{0, 0\}, \{2, 2\}, d=3-2\ep]\,, \notag\\
 & k_sI[\{1, 1, 1, 1\}, \{0, 0\}, \{2, 2\}, d=3-2\ep]\,.
\end{align}
	\item[(d)] Another logarithmic pairing is obtained by placing the
$P_2$-connection along the $x_1$ fiber and pairing it with
$\di\log x_2$, where we have $\Phi_{\bm m\bm n}\propto
(\partial_{P_2}\Omega_{ex})_{\bm m\bm n}\di x_1
\frac{\di x_2}{x_2}$ and
\begin{align}
	& k_sI[\{0, 1, 0, 0\}, \{0, 1\}, \{1, 2\}, d=3-2\ep]\,, \notag\\
	&k_sI[\{0, 1, 1, 1\}, \{0, 1\}, \{1, 2\}, d=3-2\ep]\,, \nonumber\\
	& k_sI[\{1, 0, 0, 0\}, \{0, 1\}, \{1, 2\}, d=3-2\ep]\,, \notag\\
	&k_sI[\{1, 0, 1, 1\}, \{0, 1\}, \{1, 2\}, d=3-2\ep]\,.
\end{align}
\item[(e)] Finally, the choice
\(
\Phi_{\bm m\bm n}\propto
(\partial_{x_2}\Omega_{ex})_{\bm m\bm n}
\di x_1\,\di\log x_2
\)
requires an additional \(x_2\)-IBP. The IBP generates a separately
logarithmic Gram term; after subtracting it, the two independent
representatives in the original family are
\begin{align}
 &(2+b_0+2\nu)k_s
 I[\{0,1,0,1\},\{0,0\},\{1,3\},d=3-2\ep]\,,\notag\\
 &(2+b_0+2\nu)k_s
 I[\{0,1,1,0\},\{0,0\},\{1,3\},d=3-2\ep]\,.
\end{align}
The IBP relation and the Gram subtraction are detailed in the
Supplemental Material.
\end{enumerate}
The alternative pairing of the Gram form in class (a) with
$(\partial_{x_2}\Omega_{ex})\di x_2$ reduces to class (a) after
the analogous IBP and Gram subtraction, and therefore gives no
additional representative.
After imposing the symmetry relations and parity constraints, the
five classes yield \(2+3+3+4+2=14\) independent representatives,
in agreement with the top-sector IBP count for the even-even subsystem. The
\(d=1-2\ep\) representatives in classes (a) and (b) can first be
shifted to \(d=3-2\ep\) using the dimensional recurrence relation
\eqref{shift}, after which the IBP reduction is performed uniformly.

The homogeneous \(14\times14\) top-to-top block of the differential
equations for the basis above is verified to be of
\(\di\log\)-form. The corresponding alphabet is
\begin{align}
\mathcal A=\{x-1,x,x+1\},
\end{align}
with \(x=k_s/P_1=k_s/P_2\).
The $\di\log$ decompositions of both this homogeneous block and
the complete $19\times19$ connection, including the
remaining-term family, are provided in the Supplemental Material. The three letters have clear physical intuition: \(x\), \(x+1\), and \(x-1\) correspond to the soft limit $k_s=0$, the partial-energy pole $P_{1,2}+k_s=0$, and the spurious folded pole $P_{1,2}-k_s=0$, respectively, while the total-energy pole \(P_1+P_2=0\)
corresponds to \(x\to\infty\), where the differential
system has its residue at infinity.

\section{Conclusion}
In this Letter, we study IBP reduction and differential equations for the one-loop bubble family in dS space. We first identify the parity structure of the dS IBP system, which reduces the size of the IBP system for an \(n\)-propagator family by a factor of \(2^n\), and thus significantly enhances the feasibility of this computational approach. We successfully apply the IBP reduction and the method of differential equations to loop-level massive cosmological correlators. This demonstrates the viability of these methods as a systematic and automated framework for perturbative field theory calculations in dS space.
 We also explore the interesting mathematical structure of this system. We further propose that fibration intersection theory can be generalized to cases in which the twist $\mathbf U$ is not of polynomial type. Based on this, we conjecture that selecting a $\di\log$-form twist connection and choosing $\di\log$-form representatives $\Phi$ for the master integrals can lead directly to $\di\log$-form differential equations. We adapt the Baikov representation to the dS case and find it useful for the construction of \(\di\log\)-form \(\Phi\). The results for the bubble example verify the conjecture and yield a simple alphabet. This structure may be useful for future analytic evaluations of the system.

\begin{acknowledgments}
We thank Yiyang Zhang for valuable discussions on the construction of master integrals. Jiaqi Chen is supported by the National Natural Science Foundation of China (NSFC) through Grants No. 12505094 and Science Foundation of China University of Petroleum, Beijing (No.2462025YJRC019).	
Bo Feng is supported by the NSFC through Grants No. 12535003, No.11935013, No.11947301, No.12047502. Yi-Xiao Tao is supported by the NSFC through Grant No. 124B2094. Zhehan Qin is supported by the NSFC through Grant No.\ 12275146, the National Key R\&D Program of China (2021YFC2203100), and the Dushi Program of Tsinghua University.
\end{acknowledgments}

\bibliographystyle{apsrev4-1}
\bibliography{letter}

@article{Werth:2024aui,
    author = "Werth, Denis and Pinol, Lucas and Renaux-Petel, S{\'e}bastien",
    title = "{CosmoFlow: Python Package for Cosmological Correlators}",
    eprint = "2402.03693",
    archivePrefix = "arXiv",
    primaryClass = "astro-ph.CO",
    doi = "10.1088/1361-6382/ad6740",
    journal = "Class. Quant. Grav.",
    volume = "41",
    number = "17",
    pages = "175015",
    year = "2024"
}

@article{Pinol:2023oux,
    author = "Pinol, Lucas and Renaux-Petel, S{\'e}bastien and Werth, Denis",
    title = "{The cosmological flow: a systematic approach to primordial correlators}",
    eprint = "2312.06559",
    archivePrefix = "arXiv",
    primaryClass = "astro-ph.CO",
    doi = "10.1088/1475-7516/2025/02/019",
    journal = "JCAP",
    volume = "02",
    pages = "019",
    year = "2025"
}

@article{Werth:2023pfl,
    author = "Werth, Denis and Pinol, Lucas and Renaux-Petel, S{\'e}bastien",
    title = "{Cosmological Flow of Primordial Correlators}",
    eprint = "2302.00655",
    archivePrefix = "arXiv",
    primaryClass = "hep-th",
    doi = "10.1103/PhysRevLett.133.141002",
    journal = "Phys. Rev. Lett.",
    volume = "133",
    number = "14",
    pages = "141002",
    year = "2024"
}

@article{Wang:2021qez,
    author = "Wang, Lian-Tao and Xianyu, Zhong-Zhi and Zhong, Yi-Ming",
    title = "{Precision calculation of inflation correlators at one loop}",
    eprint = "2109.14635",
    archivePrefix = "arXiv",
    primaryClass = "hep-ph",
    doi = "10.1007/JHEP02(2022)085",
    journal = "JHEP",
    volume = "02",
    pages = "085",
    year = "2022"
}

@article{Das:2026vfv,
    author = "Das, Shibam and Karan, Debanjan and Khatun, Babli and Kundu, Nilay",
    title = "{Reconstructing cosmological correlators via dispersion: from cutting to dressing rules}",
    eprint = "2602.05546",
    archivePrefix = "arXiv",
    primaryClass = "hep-th",
    month = "2",
    year = "2026"
}

@article{Qin:2025xct,
	archiveprefix = {arXiv},
	author = {Qin, Zhehan and Renaux-Petel, S{\'e}bastien and Tong, Xi and Werth, Denis and Zhu, Yuhang},
	doi = {10.1088/1475-7516/2025/09/058},
	eprint = {2506.01555},
	journal = {JCAP},
	pages = {058},
	primaryclass = {hep-th},
	title = {{The exact and approximate tales of boost-breaking cosmological correlators}},
	volume = {09},
	year = {2025}}

@article{Cespedes:2025ple,
	archiveprefix = {arXiv},
	author = {Cespedes, Sebastian and Qin, Zhehan and Wang, Dong-Gang},
	eprint = {2510.25826},
	month = {10},
	primaryclass = {hep-th},
	reportnumber = {Imperial-TP-2025-SCC-3},
	title = {{$\lambda \phi^4${\textasciitilde}as an Effective Theory in de Sitter}},
	year = {2025}}

@article{Fan:2024iek,
	archiveprefix = {arXiv},
	author = {Fan, Bingchu and Xianyu, Zhong-Zhi},
	doi = {10.1007/JHEP12(2024)042},
	eprint = {2403.07050},
	journal = {JHEP},
	pages = {042},
	primaryclass = {hep-th},
	title = {{Cosmological amplitudes in power-law FRW universe}},
	volume = {12},
	year = {2024}}

@article{Bern:1993kr,
	archiveprefix = {arXiv},
	author = {Bern, Zvi and Dixon, Lance J. and Kosower, David A.},
	doi = {10.1016/0550-3213(94)90398-0},
	eprint = {hep-ph/9306240},
	journal = {Nucl. Phys. B},
	pages = {751--816},
	reportnumber = {SLAC-PUB-5947, SACLAY-SPH-T-92-048, UCLA-92-043},
	title = {{Dimensionally regulated pentagon integrals}},
	volume = {412},
	year = {1994}}

@article{Gehrmann:1999as,
	archiveprefix = {arXiv},
	author = {Gehrmann, T. and Remiddi, E.},
	doi = {10.1016/S0550-3213(00)00223-6},
	eprint = {hep-ph/9912329},
	journal = {Nucl. Phys. B},
	pages = {485--518},
	reportnumber = {TTP-99-49},
	title = {{Differential equations for two-loop four-point functions}},
	volume = {580},
	year = {2000}}

@article{Kotikov:1991pm,
	author = {Kotikov, A. V.},
	doi = {10.1016/0370-2693(91)90536-Y},
	journal = {Phys. Lett. B},
	note = {[Erratum: Phys.Lett.B 295, 409--409 (1992)]},
	pages = {123--127},
	title = {{Differential equation method: The Calculation of N point Feynman diagrams}},
	volume = {267},
	year = {1991}}

@article{Kotikov:1990kg,
	author = {Kotikov, A. V.},
	doi = {10.1016/0370-2693(91)90413-K},
	journal = {Phys. Lett. B},
	pages = {158--164},
	reportnumber = {ITF-90-31E},
	title = {{Differential equations method: New technique for massive Feynman diagrams calculation}},
	volume = {254},
	year = {1991}}

@article{Chetyrkin:1981qh,
	author = {Chetyrkin, K. G. and Tkachov, F. V.},
	doi = {10.1016/0550-3213(81)90199-1},
	journal = {Nucl. Phys. B},
	pages = {159--204},
	title = {{Integration by parts: The algorithm to calculate $\beta$-functions in 4 loops}},
	volume = {192},
	year = {1981}}

@article{Qin:2023bjk,
	archiveprefix = {arXiv},
	author = {Qin, Zhehan and Xianyu, Zhong-Zhi},
	doi = {10.1007/JHEP09(2023)116},
	eprint = {2304.13295},
	journal = {JHEP},
	pages = {116},
	primaryclass = {hep-th},
	title = {{Inflation correlators at the one-loop order: nonanalyticity, factorization, cutting rule, and OPE}},
	volume = {09},
	year = {2023}}

@article{Wang:2025qfh,
	archiveprefix = {arXiv},
	author = {Wang, Donggang and Wang, Xiangwei and Wang, Yi and Yu, Wenqi},
	eprint = {2508.12856},
	month = {8},
	primaryclass = {hep-th},
	title = {{Interact or Twist: Cosmological Correlators from Field Redefinitions Revisited}},
	year = {2025}}

@article{Pimentel:2026kqc,
	archiveprefix = {arXiv},
	author = {Pimentel, Guilherme L. and Westerdijk, Tom},
	eprint = {2601.00952},
	month = {1},
	primaryclass = {hep-th},
	title = {{On Cosmological Correlators at One Loop}},
	year = {2026}}

@article{Brunello:2024tqf,
	archiveprefix = {arXiv},
	author = {Brunello, Giacomo and Chestnov, Vsevolod and Mastrolia, Pierpaolo},
	doi = {10.1007/JHEP07(2025)045},
	eprint = {2408.16668},
	journal = {JHEP},
	pages = {045},
	primaryclass = {hep-th},
	title = {{Intersection numbers from companion tensor algebra}},
	volume = {07},
	year = {2025}}

@article{Brunello:2023rpq,
	archiveprefix = {arXiv},
	author = {Brunello, Giacomo and Chestnov, Vsevolod and Crisanti, Giulio and Frellesvig, Hjalte and Mandal, Manoj K. and Mastrolia, Pierpaolo},
	doi = {10.1007/JHEP09(2024)015},
	eprint = {2401.01897},
	journal = {JHEP},
	pages = {015},
	primaryclass = {hep-th},
	title = {{Intersection numbers, polynomial division and relative cohomology}},
	volume = {09},
	year = {2024}}

@article{Fontana:2023amt,
	archiveprefix = {arXiv},
	author = {Fontana, Gaia and Peraro, Tiziano},
	doi = {10.1007/JHEP08(2023)175},
	eprint = {2304.14336},
	journal = {JHEP},
	pages = {175},
	primaryclass = {hep-ph},
	reportnumber = {ZU-TH 19/23},
	title = {{Reduction to master integrals via intersection numbers and polynomial expansions}},
	volume = {08},
	year = {2023}}

@article{Duhr:2024rxe,
	archiveprefix = {arXiv},
	author = {Duhr, Claude and Porkert, Franziska and Semper, Cathrin and Stawinski, Sven F.},
	doi = {10.1007/JHEP03(2025)019},
	eprint = {2407.17175},
	journal = {JHEP},
	pages = {019},
	primaryclass = {hep-th},
	reportnumber = {BONN-TH-2024-10},
	title = {{Twisted Riemann bilinear relations and Feynman integrals}},
	volume = {03},
	year = {2025}}

@article{Giroux:2022wav,
	archiveprefix = {arXiv},
	author = {Giroux, Mathieu and Pokraka, Andrzej},
	doi = {10.1007/JHEP03(2023)155},
	eprint = {2210.09898},
	journal = {JHEP},
	pages = {155},
	primaryclass = {hep-th},
	title = {{Loop-by-loop differential equations for dual (elliptic) Feynman integrals}},
	volume = {03},
	year = {2023}}

@article{Caron-Huot:2021iev,
	archiveprefix = {arXiv},
	author = {Caron-Huot, Simon and Pokraka, Andrzej},
	doi = {10.1007/JHEP04(2022)078},
	eprint = {2112.00055},
	journal = {JHEP},
	pages = {078},
	primaryclass = {hep-th},
	title = {{Duals of Feynman Integrals. Part II. Generalized unitarity}},
	volume = {04},
	year = {2022}}

@article{Caron-Huot:2021xqj,
	archiveprefix = {arXiv},
	author = {Caron-Huot, Simon and Pokraka, Andrzej},
	doi = {10.1007/JHEP12(2021)045},
	eprint = {2104.06898},
	journal = {JHEP},
	pages = {045},
	primaryclass = {hep-th},
	title = {{Duals of Feynman integrals. Part I. Differential equations}},
	volume = {12},
	year = {2021}}

@article{Chestnov:2022alh,
	archiveprefix = {arXiv},
	author = {Chestnov, Vsevolod and Gasparotto, Federico and Mandal, Manoj K. and Mastrolia, Pierpaolo and Matsubara-Heo, Saiei J. and Munch, Henrik J. and Takayama, Nobuki},
	doi = {10.1007/JHEP09(2022)187},
	eprint = {2204.12983},
	journal = {JHEP},
	pages = {187},
	primaryclass = {hep-th},
	title = {{Macaulay matrix for Feynman integrals: linear relations and intersection numbers}},
	volume = {09},
	year = {2022}}

@article{Frellesvig:2020qot,
	archiveprefix = {arXiv},
	author = {Frellesvig, Hjalte and Gasparotto, Federico and Laporta, Stefano and Mandal, Manoj K. and Mastrolia, Pierpaolo and Mattiazzi, Luca and Mizera, Sebastian},
	doi = {10.1007/JHEP03(2021)027},
	eprint = {2008.04823},
	journal = {JHEP},
	pages = {027},
	primaryclass = {hep-th},
	title = {{Decomposition of Feynman Integrals by Multivariate Intersection Numbers}},
	volume = {03},
	year = {2021}}

@article{Crisanti:2024onv,
	archiveprefix = {arXiv},
	author = {Crisanti, Giulio and Smith, Sid},
	doi = {10.1007/JHEP09(2024)018},
	eprint = {2405.18178},
	journal = {JHEP},
	pages = {018},
	primaryclass = {hep-th},
	reportnumber = {MPP-2024-104},
	title = {{Feynman integral reductions by intersection theory with orthogonal bases and closed formulae}},
	volume = {09},
	year = {2024}}

@article{Jiang:2023oyq,
	archiveprefix = {arXiv},
	author = {Jiang, Xuhang and Lian, Ming and Yang, Li Lin},
	doi = {10.1103/PhysRevD.109.076020},
	eprint = {2312.03453},
	journal = {Phys. Rev. D},
	number = {7},
	pages = {076020},
	primaryclass = {hep-ph},
	title = {{Recursive structure of Baikov representations: The top-down reduction with intersection theory}},
	volume = {109},
	year = {2024}}

@article{Weinzierl:2020xyy,
	archiveprefix = {arXiv},
	author = {Weinzierl, Stefan},
	doi = {10.1063/5.0054292},
	eprint = {2002.01930},
	journal = {J. Math. Phys.},
	number = {7},
	pages = {072301},
	primaryclass = {math-ph},
	title = {{On the computation of intersection numbers for twisted cocycles}},
	volume = {62},
	year = {2021}}

@article{Chestnov:2022xsy,
	archiveprefix = {arXiv},
	author = {Chestnov, Vsevolod and Frellesvig, Hjalte and Gasparotto, Federico and Mandal, Manoj K. and Mastrolia, Pierpaolo},
	doi = {10.1007/JHEP06(2023)131},
	eprint = {2209.01997},
	journal = {JHEP},
	pages = {131},
	primaryclass = {hep-th},
	title = {{Intersection numbers from higher-order partial differential equations}},
	volume = {06},
	year = {2023}}

@article{Mizera:2019vvs,
	archiveprefix = {arXiv},
	author = {Mizera, Sebastian and Pokraka, Andrzej},
	doi = {10.1007/JHEP02(2020)159},
	eprint = {1910.11852},
	journal = {JHEP},
	pages = {159},
	primaryclass = {hep-th},
	title = {{From Infinity to Four Dimensions: Higher Residue Pairings and Feynman Integrals}},
	volume = {02},
	year = {2020}}

@article{Baumann:2011nk,
	archiveprefix = {arXiv},
	author = {Baumann, Daniel and Green, Daniel},
	doi = {10.1103/PhysRevD.85.103520},
	eprint = {1109.0292},
	journal = {Phys. Rev. D},
	pages = {103520},
	primaryclass = {hep-th},
	title = {{Signatures of Supersymmetry from the Early Universe}},
	volume = {85},
	year = {2012}}

@article{Noumi:2012vr,
	archiveprefix = {arXiv},
	author = {Noumi, Toshifumi and Yamaguchi, Masahide and Yokoyama, Daisuke},
	doi = {10.1007/JHEP06(2013)051},
	eprint = {1211.1624},
	journal = {JHEP},
	pages = {051},
	primaryclass = {hep-th},
	reportnumber = {UT-KOMABA-12-9, TIT-HEP-625},
	title = {{Effective field theory approach to quasi-single field inflation and effects of heavy fields}},
	volume = {06},
	year = {2013}}

@article{Glew:2025otn,
	archiveprefix = {arXiv},
	author = {Glew, Ross and Lukowski, Tomasz},
	doi = {10.1007/JHEP09(2025)074},
	eprint = {2502.17564},
	journal = {JHEP},
	pages = {074},
	primaryclass = {hep-th},
	title = {{Amplitubes: graph cosmohedra}},
	volume = {09},
	year = {2025}}

@article{De:2025bmf,
	archiveprefix = {arXiv},
	author = {De, Shounak and Paranjape, Shruti and Pokraka, Andrzej and Spradlin, Marcus and Volovich, Anastasia},
	doi = {10.1007/JHEP07(2025)174},
	eprint = {2503.23579},
	journal = {JHEP},
	pages = {174},
	primaryclass = {hep-th},
	title = {{Hidden zeros of the cosmological wavefunction}},
	volume = {07},
	year = {2025}}

@article{Glew:2025ypb,
	archiveprefix = {arXiv},
	author = {Glew, Ross and Pokraka, Andrzej},
	eprint = {2508.11568},
	month = {8},
	primaryclass = {hep-th},
	title = {{Kinematic flow from the flow of cuts}},
	year = {2025}}

@article{Chowdhury:2025ohm,
	archiveprefix = {arXiv},
	author = {Chowdhury, Chandramouli and Lipstein, Arthur and Marshall, Joe and Mei, Jiajie and Sachs, Ivo},
	doi = {10.1007/JHEP03(2026)076},
	eprint = {2503.10598},
	journal = {JHEP},
	pages = {076},
	primaryclass = {hep-th},
	reportnumber = {LMU-ASC 03/25},
	title = {{Cosmological dressing rules}},
	volume = {03},
	year = {2026}}

@article{Baumann:2024mvm,
	archiveprefix = {arXiv},
	author = {Baumann, Daniel and Goodhew, Harry and Lee, Hayden},
	doi = {10.1007/JHEP07(2025)131},
	eprint = {2410.17994},
	journal = {JHEP},
	pages = {131},
	primaryclass = {hep-th},
	title = {{Kinematic flow for cosmological loop integrands}},
	volume = {07},
	year = {2025}}

@article{Benincasa:2022gtd,
	archiveprefix = {arXiv},
	author = {Benincasa, Paolo},
	doi = {10.1142/S0217751X22300101},
	eprint = {2203.15330},
	month = {3},
	primaryclass = {hep-th},
	title = {{Amplitudes meet Cosmology: A (Scalar) Primer}},
	year = {2022}}

@article{Benincasa:2024lxe,
	archiveprefix = {arXiv},
	author = {Benincasa, Paolo and Vaz{\~a}o, Francisco},
	doi = {10.21468/SciPostPhys.19.2.029},
	eprint = {2402.06558},
	journal = {SciPost Phys.},
	number = {2},
	pages = {029},
	primaryclass = {hep-th},
	title = {{The asymptotic structure of cosmological integrals}},
	volume = {19},
	year = {2025}}

@article{Benincasa:2024leu,
	archiveprefix = {arXiv},
	author = {Benincasa, Paolo and Dian, Gabriele},
	doi = {10.21468/SciPostPhys.18.3.105},
	eprint = {2401.05207},
	journal = {SciPost Phys.},
	pages = {105},
	primaryclass = {hep-th},
	reportnumber = {MPP-2023-150, DESY-24-006},
	title = {{The Geometry of Cosmological Correlators}},
	volume = {18},
	year = {2025}}

@article{Arkani-Hamed:2023bsv,
	archiveprefix = {arXiv},
	author = {Arkani-Hamed, Nima and Baumann, Daniel and Hillman, Aaron and Joyce, Austin and Lee, Hayden and Pimentel, Guilherme L.},
	doi = {10.1103/dsjm-tckw},
	eprint = {2312.05300},
	journal = {Phys. Rev. Lett.},
	number = {3},
	pages = {031602},
	primaryclass = {hep-th},
	title = {{Kinematic Flow and the Emergence of Time}},
	volume = {135},
	year = {2025}}

@article{Arkani-Hamed:2024jbp,
	archiveprefix = {arXiv},
	author = {Arkani-Hamed, Nima and Figueiredo, Carolina and Vaz{\~a}o, Francisco},
	doi = {10.1007/JHEP11(2025)029},
	eprint = {2412.19881},
	journal = {JHEP},
	pages = {029},
	primaryclass = {hep-th},
	title = {{Cosmohedra}},
	volume = {11},
	year = {2025}}

@article{Xianyu:2023ytd,
	archiveprefix = {arXiv},
	author = {Xianyu, Zhong-Zhi and Zang, Jiaju},
	doi = {10.1007/JHEP03(2024)070},
	eprint = {2309.10849},
	journal = {JHEP},
	pages = {070},
	primaryclass = {hep-th},
	title = {{Inflation correlators with multiple massive exchanges}},
	volume = {03},
	year = {2024}}

@article{Qin:2022lva,
	archiveprefix = {arXiv},
	author = {Qin, Zhehan and Xianyu, Zhong-Zhi},
	doi = {10.1007/JHEP10(2022)192},
	eprint = {2205.01692},
	journal = {JHEP},
	pages = {192},
	primaryclass = {hep-th},
	title = {{Phase information in cosmological collider signals}},
	volume = {10},
	year = {2022}}

@article{Altshuler:2025qmk,
	archiveprefix = {arXiv},
	author = {Altshuler, Boris L.},
	doi = {10.1007/JHEP12(2025)102},
	eprint = {2508.07467},
	journal = {JHEP},
	pages = {102},
	primaryclass = {hep-th},
	title = {{Kallen-Lehmann representation for Spinor-Scalar Loops in de Sitter Space-time. Spectral equations}},
	volume = {12},
	year = {2025}}

@article{Zhang:2025nzd,
	archiveprefix = {arXiv},
	author = {Zhang, Hongyu},
	doi = {10.1007/JHEP02(2026)119},
	eprint = {2507.19318},
	journal = {JHEP},
	pages = {119},
	primaryclass = {hep-th},
	title = {{Dimensional regularization of bubble diagrams in de Sitter spacetime}},
	volume = {02},
	year = {2026}}

@article{Werth:2024mjg,
	archiveprefix = {arXiv},
	author = {Werth, Denis},
	doi = {10.1007/JHEP12(2024)017},
	eprint = {2409.02072},
	journal = {JHEP},
	pages = {017},
	primaryclass = {hep-th},
	title = {{Spectral representation of cosmological correlators}},
	volume = {12},
	year = {2024}}

@article{Loparco:2023rug,
	archiveprefix = {arXiv},
	author = {Loparco, Manuel and Penedones, Joao and Salehi Vaziri, Kamran and Sun, Zimo},
	doi = {10.1007/JHEP12(2023)159},
	eprint = {2306.00090},
	journal = {JHEP},
	pages = {159},
	primaryclass = {hep-th},
	title = {{The K{\"a}ll{\'e}n-Lehmann representation in de Sitter spacetime}},
	volume = {12},
	year = {2023}}

@article{Marolf:2010zp,
	archiveprefix = {arXiv},
	author = {Marolf, Donald and Morrison, Ian A.},
	doi = {10.1103/PhysRevD.82.105032},
	eprint = {1006.0035},
	journal = {Phys. Rev. D},
	pages = {105032},
	primaryclass = {gr-qc},
	title = {{The IR stability of de Sitter: Loop corrections to scalar propagators}},
	volume = {82},
	year = {2010}}

@article{Liu:2024xyi,
	archiveprefix = {arXiv},
	author = {Liu, Haoyuan and Qin, Zhehan and Xianyu, Zhong-Zhi},
	doi = {10.1007/JHEP02(2025)101},
	eprint = {2407.12299},
	journal = {JHEP},
	pages = {101},
	primaryclass = {hep-th},
	reportnumber = {USTC-ICTS/PCFT-24-23},
	title = {{Dispersive bootstrap of massive inflation correlators}},
	volume = {02},
	year = {2025}}

@article{Qin:2024gtr,
	archiveprefix = {arXiv},
	author = {Qin, Zhehan},
	doi = {10.1007/JHEP03(2025)051},
	eprint = {2411.13636},
	journal = {JHEP},
	pages = {051},
	primaryclass = {hep-th},
	title = {{Cosmological correlators at the loop level}},
	volume = {03},
	year = {2025}}

@article{Premkumar:2021mlz,
	archiveprefix = {arXiv},
	author = {Premkumar, Akhil},
	doi = {10.1103/PhysRevD.109.045003},
	eprint = {2110.12504},
	journal = {Phys. Rev. D},
	number = {4},
	pages = {045003},
	primaryclass = {hep-th},
	title = {{Regulating loops in de Sitter spacetime}},
	volume = {109},
	year = {2024}}

@article{Sleight:2020obc,
	archiveprefix = {arXiv},
	author = {Sleight, Charlotte and Taronna, Massimo},
	doi = {10.1103/PhysRevD.104.L081902},
	eprint = {2007.09993},
	journal = {Phys. Rev. D},
	number = {8},
	pages = {L081902},
	primaryclass = {hep-th},
	title = {{From AdS to dS exchanges: Spectral representation, Mellin amplitudes, and crossing}},
	volume = {104},
	year = {2021}}

@article{Sleight:2019hfp,
	archiveprefix = {arXiv},
	author = {Sleight, Charlotte and Taronna, Massimo},
	doi = {10.1007/JHEP02(2020)098},
	eprint = {1907.01143},
	journal = {JHEP},
	pages = {098},
	primaryclass = {hep-th},
	reportnumber = {PUPT-2590},
	title = {{Bootstrapping Inflationary Correlators in Mellin Space}},
	volume = {02},
	year = {2020}}

@article{Sleight:2019mgd,
	archiveprefix = {arXiv},
	author = {Sleight, Charlotte},
	doi = {10.1007/JHEP01(2020)090},
	eprint = {1906.12302},
	journal = {JHEP},
	pages = {090},
	primaryclass = {hep-th},
	title = {{A Mellin Space Approach to Cosmological Correlators}},
	volume = {01},
	year = {2020}}

@article{Xianyu:2025lbk,
	archiveprefix = {arXiv},
	author = {Xianyu, Zhong-Zhi and Zang, Jiaju},
	doi = {10.1007/JHEP03(2026)122},
	eprint = {2511.08677},
	journal = {JHEP},
	pages = {122},
	primaryclass = {hep-th},
	title = {{Massive inflationary amplitudes: new representations and degenerate limits}},
	volume = {03},
	year = {2026}}

@article{Liu:2024str,
	archiveprefix = {arXiv},
	author = {Liu, Haoyuan and Xianyu, Zhong-Zhi},
	doi = {10.1007/JHEP09(2025)183},
	eprint = {2412.07843},
	journal = {JHEP},
	pages = {183},
	primaryclass = {hep-th},
	reportnumber = {USTC-ICTS/PCFT-24-56},
	title = {{Massive inflationary amplitudes: differential equations and complete solutions for general trees}},
	volume = {09},
	year = {2025}}

@article{Aoki:2024uyi,
	archiveprefix = {arXiv},
	author = {Aoki, Shuntaro and Pinol, Lucas and Sano, Fumiya and Yamaguchi, Masahide and Zhu, Yuhang},
	doi = {10.1007/JHEP09(2024)176},
	eprint = {2404.09547},
	journal = {JHEP},
	pages = {176},
	primaryclass = {hep-th},
	title = {{Cosmological correlators with double massive exchanges: bootstrap equation and phenomenology}},
	volume = {09},
	year = {2024}}

@article{Qin:2023ejc,
	archiveprefix = {arXiv},
	author = {Qin, Zhehan and Xianyu, Zhong-Zhi},
	doi = {10.1007/JHEP07(2023)001},
	eprint = {2301.07047},
	journal = {JHEP},
	pages = {001},
	primaryclass = {hep-th},
	title = {{Closed-form formulae for inflation correlators}},
	volume = {07},
	year = {2023}}

@article{Wang:2022eop,
	archiveprefix = {arXiv},
	author = {Wang, Dong-Gang and Pimentel, Guilherme L. and Ach{\'u}carro, Ana},
	doi = {10.1088/1475-7516/2023/05/043},
	eprint = {2212.14035},
	journal = {JCAP},
	pages = {043},
	primaryclass = {astro-ph.CO},
	title = {{Bootstrapping multi-field inflation: non-Gaussianities from light scalars revisited}},
	volume = {05},
	year = {2023}}

@article{Qin:2022fbv,
	archiveprefix = {arXiv},
	author = {Qin, Zhehan and Xianyu, Zhong-Zhi},
	doi = {10.1007/JHEP04(2023)059},
	eprint = {2208.13790},
	journal = {JHEP},
	pages = {059},
	primaryclass = {hep-th},
	title = {{Helical inflation correlators: partial Mellin-Barnes and bootstrap equations}},
	volume = {04},
	year = {2023}}

@article{Jazayeri:2022kjy,
	archiveprefix = {arXiv},
	author = {Jazayeri, Sadra and Renaux-Petel, S{\'e}bastien},
	doi = {10.1007/JHEP12(2022)137},
	eprint = {2205.10340},
	journal = {JHEP},
	pages = {137},
	primaryclass = {hep-th},
	title = {{Cosmological bootstrap in slow motion}},
	volume = {12},
	year = {2022}}

@article{Pimentel:2022fsc,
	archiveprefix = {arXiv},
	author = {Pimentel, Guilherme L. and Wang, Dong-Gang},
	doi = {10.1007/JHEP10(2022)177},
	eprint = {2205.00013},
	journal = {JHEP},
	pages = {177},
	primaryclass = {hep-th},
	title = {{Boostless cosmological collider bootstrap}},
	volume = {10},
	year = {2022}}

@article{Baumann:2022jpr,
	archiveprefix = {arXiv},
	author = {Baumann, Daniel and Green, Daniel and Joyce, Austin and Pajer, Enrico and Pimentel, Guilherme L. and Sleight, Charlotte and Taronna, Massimo},
	doi = {10.21468/SciPostPhysCommRep.1},
	eprint = {2203.08121},
	journal = {SciPost Phys. Comm. Rep.},
	pages = {1},
	primaryclass = {hep-th},
	title = {{Snowmass White Paper: The Cosmological Bootstrap}},
	volume = {2024},
	year = {2024}}

@article{Cabass:2021fnw,
	archiveprefix = {arXiv},
	author = {Cabass, Giovanni and Pajer, Enrico and Stefanyszyn, David and Supe{\l}, Jakub},
	doi = {10.1007/JHEP05(2022)077},
	eprint = {2109.10189},
	journal = {JHEP},
	pages = {077},
	primaryclass = {hep-th},
	title = {{Bootstrapping large graviton non-Gaussianities}},
	volume = {05},
	year = {2022}}

@article{Pajer:2020wxk,
	archiveprefix = {arXiv},
	author = {Pajer, Enrico},
	doi = {10.1088/1475-7516/2021/01/023},
	eprint = {2010.12818},
	journal = {JCAP},
	pages = {023},
	primaryclass = {hep-th},
	title = {{Building a Boostless Bootstrap for the Bispectrum}},
	volume = {01},
	year = {2021}}

@article{Pajer:2020wnj,
	archiveprefix = {arXiv},
	author = {Pajer, Enrico and Stefanyszyn, David and Supe{\l}, Jakub},
	doi = {10.1007/JHEP12(2020)198},
	eprint = {2007.00027},
	journal = {JHEP},
	note = {[Erratum: JHEP 04, 023 (2022)]},
	pages = {198},
	primaryclass = {hep-th},
	title = {{The Boostless Bootstrap: Amplitudes without Lorentz boosts}},
	volume = {12},
	year = {2020}}

@article{Baumann:2020dch,
	archiveprefix = {arXiv},
	author = {Baumann, Daniel and Duaso Pueyo, Carlos and Joyce, Austin and Lee, Hayden and Pimentel, Guilherme L.},
	doi = {10.21468/SciPostPhys.11.3.071},
	eprint = {2005.04234},
	journal = {SciPost Phys.},
	pages = {071},
	primaryclass = {hep-th},
	title = {{The Cosmological Bootstrap: Spinning Correlators from Symmetries and Factorization}},
	volume = {11},
	year = {2021}}

@article{Achucarro:2022qrl,
	archiveprefix = {arXiv},
	author = {Ach{\'u}carro, Ana and others},
	eprint = {2203.08128},
	month = {3},
	primaryclass = {astro-ph.CO},
	title = {{Inflation: Theory and Observations}},
	year = {2022}}

@article{Chen:2024glu,
	archiveprefix = {arXiv},
	author = {Chen, Jiaqi and Feng, Bo and Tao, Yi-Xiao},
	doi = {10.1007/JHEP03(2025)075},
	eprint = {2411.03088},
	journal = {JHEP},
	pages = {075},
	primaryclass = {hep-th},
	title = {{Multivariate hypergeometric solutions of cosmological (dS) correlators by d log-form differential equations}},
	volume = {03},
	year = {2025}}

@article{Lee:2010wea,
	archiveprefix = {arXiv},
	author = {Lee, R. N.},
	doi = {10.1016/j.nuclphysbps.2010.08.032},
	editor = {Bl{\"u}mlein, Johannes and Moch, Sven-Olaf and Riemann, Tord},
	eprint = {1007.2256},
	journal = {Nucl. Phys. B Proc. Suppl.},
	pages = {135--140},
	primaryclass = {hep-ph},
	title = {{Calculating multiloop integrals using dimensional recurrence relation and $D$-analyticity}},
	volume = {205-206},
	year = {2010}}

@article{Tarasov:1996br,
	archiveprefix = {arXiv},
	author = {Tarasov, O. V.},
	doi = {10.1103/PhysRevD.54.6479},
	eprint = {hep-th/9606018},
	journal = {Phys. Rev. D},
	pages = {6479--6490},
	reportnumber = {DESY-96-068, JINR-E2-96-62},
	title = {{Connection between Feynman integrals having different values of the space-time dimension}},
	volume = {54},
	year = {1996}}

@article{Baikov:1996iu,
	archiveprefix = {arXiv},
	author = {Baikov, P. A.},
	doi = {10.1016/S0168-9002(97)00126-5},
	editor = {Werlen, M. and Perret-Gallix, D.},
	eprint = {hep-ph/9611449},
	journal = {Nucl. Instrum. Meth. A},
	pages = {347--349},
	reportnumber = {INP-96-42-449},
	title = {{Explicit solutions of the multiloop integral recurrence relations and its application}},
	volume = {389},
	year = {1997}}

@article{Lange:2025fba,
	archiveprefix = {arXiv},
	author = {Lange, Fabian and Usovitsch, Johann and Wu, Zihao},
	doi = {10.1016/j.cpc.2025.109999},
	eprint = {2505.20197},
	journal = {Comput. Phys. Commun.},
	pages = {109999},
	primaryclass = {hep-ph},
	reportnumber = {ZU-TH 39/25, HU-EP-25/17-RTG},
	title = {{Kira 3: integral reduction with efficient seeding and optimized equation selection}},
	volume = {322},
	year = {2026}}

@article{Klappert:2020nbg,
	archiveprefix = {arXiv},
	author = {Klappert, Jonas and Lange, Fabian and Maierh{\"o}fer, Philipp and Usovitsch, Johann},
	doi = {10.1016/j.cpc.2021.108024},
	eprint = {2008.06494},
	journal = {Comput. Phys. Commun.},
	pages = {108024},
	primaryclass = {hep-ph},
	reportnumber = {TTK-20-24, P3H-20-041, FR-PHENO-2020-11, MITP/20-044},
	title = {{Integral reduction with Kira 2.0 and finite field methods}},
	volume = {266},
	year = {2021}}

@article{Maierhofer:2017gsa,
	archiveprefix = {arXiv},
	author = {Maierh{\"o}fer, Philipp and Usovitsch, Johann and Uwer, Peter},
	doi = {10.1016/j.cpc.2018.04.012},
	eprint = {1705.05610},
	journal = {Comput. Phys. Commun.},
	pages = {99--112},
	primaryclass = {hep-ph},
	title = {{Kira{\textemdash}A Feynman integral reduction program}},
	volume = {230},
	year = {2018}}

@article{Chen:2023iix,
	archiveprefix = {arXiv},
	author = {Chen, Jiaqi and Feng, Bo},
	doi = {10.1007/JHEP06(2024)199},
	eprint = {2401.00129},
	journal = {JHEP},
	pages = {199},
	primaryclass = {hep-th},
	title = {{Towards systematic evaluation of de Sitter correlators via Generalized Integration-By-Parts relations}},
	volume = {06},
	year = {2024}}

@article{Chen:2017ryl,
	archiveprefix = {arXiv},
	author = {Chen, Xingang and Wang, Yi and Xianyu, Zhong-Zhi},
	doi = {10.1088/1475-7516/2017/12/006},
	eprint = {1703.10166},
	journal = {JCAP},
	pages = {006},
	primaryclass = {hep-th},
	title = {{Schwinger-Keldysh Diagrammatics for Primordial Perturbations}},
	volume = {12},
	year = {2017}}

@article{Schwinger:1960qe,
	author = {Schwinger, Julian S.},
	doi = {10.1063/1.1703727},
	journal = {J. Math. Phys.},
	pages = {407--432},
	title = {{Brownian motion of a quantum oscillator}},
	volume = {2},
	year = {1961}}

@article{Feynman:1963fq,
	author = {Feynman, R. P. and Vernon, Jr., F. L.},
	doi = {10.1016/0003-4916(63)90068-X},
	editor = {Brown, L. M.},
	journal = {Annals Phys.},
	pages = {118--173},
	title = {{The Theory of a general quantum system interacting with a linear dissipative system}},
	volume = {24},
	year = {1963}}

@article{Keldysh:1964ud,
	author = {Keldysh, L. V.},
	doi = {10.1142/9789811279461_0007},
	journal = {Sov. Phys. JETP},
	pages = {1018--1026},
	title = {{Diagram Technique for Nonequilibrium Processes}},
	volume = {20},
	year = {1965}}

@article{Weinberg:2005vy,
	archiveprefix = {arXiv},
	author = {Weinberg, Steven},
	doi = {10.1103/PhysRevD.72.043514},
	eprint = {hep-th/0506236},
	journal = {Phys. Rev. D},
	pages = {043514},
	reportnumber = {UTTG-01-05},
	title = {{Quantum contributions to cosmological correlations}},
	volume = {72},
	year = {2005}}

@article{Chen:2009we,
	archiveprefix = {arXiv},
	author = {Chen, Xingang and Wang, Yi},
	doi = {10.1103/PhysRevD.81.063511},
	eprint = {0909.0496},
	journal = {Phys. Rev. D},
	pages = {063511},
	primaryclass = {astro-ph.CO},
	reportnumber = {MIT-CTP-4071},
	title = {{Large non-Gaussianities with Intermediate Shapes from Quasi-Single Field Inflation}},
	volume = {81},
	year = {2010}}

@article{Chen:2009zp,
	archiveprefix = {arXiv},
	author = {Chen, Xingang and Wang, Yi},
	doi = {10.1088/1475-7516/2010/04/027},
	eprint = {0911.3380},
	journal = {JCAP},
	pages = {027},
	primaryclass = {hep-th},
	title = {{Quasi-Single Field Inflation and Non-Gaussianities}},
	volume = {04},
	year = {2010}}

@article{Arkani-Hamed:2015bza,
	archiveprefix = {arXiv},
	author = {Arkani-Hamed, Nima and Maldacena, Juan},
	eprint = {1503.08043},
	month = {3},
	primaryclass = {hep-th},
	title = {{Cosmological Collider Physics}},
	year = {2015}}

@article{Arkani-Hamed:2018kmz,
	archiveprefix = {arXiv},
	author = {Arkani-Hamed, Nima and Baumann, Daniel and Lee, Hayden and Pimentel, Guilherme L.},
	doi = {10.1007/JHEP04(2020)105},
	eprint = {1811.00024},
	journal = {JHEP},
	pages = {105},
	primaryclass = {hep-th},
	title = {{The Cosmological Bootstrap: Inflationary Correlators from Symmetries and Singularities}},
	volume = {04},
	year = {2020}}

@article{Baumann:2019oyu,
	archiveprefix = {arXiv},
	author = {Baumann, Daniel and Duaso Pueyo, Carlos and Joyce, Austin and Lee, Hayden and Pimentel, Guilherme L.},
	doi = {10.1007/JHEP12(2020)204},
	eprint = {1910.14051},
	journal = {JHEP},
	pages = {204},
	primaryclass = {hep-th},
	title = {{The cosmological bootstrap: weight-shifting operators and scalar seeds}},
	volume = {12},
	year = {2020}}

@article{Sleight:2021plv,
	archiveprefix = {arXiv},
	author = {Sleight, Charlotte and Taronna, Massimo},
	doi = {10.1007/JHEP12(2021)074},
	eprint = {2109.02725},
	journal = {JHEP},
	pages = {074},
	primaryclass = {hep-th},
	title = {{From dS to AdS and back}},
	volume = {12},
	year = {2021}}

@article{Xianyu:2022jwk,
	archiveprefix = {arXiv},
	author = {Xianyu, Zhong-Zhi and Zhang, Hongyu},
	doi = {10.1007/JHEP04(2023)103},
	eprint = {2211.03810},
	journal = {JHEP},
	pages = {103},
	primaryclass = {hep-th},
	title = {{Bootstrapping one-loop inflation correlators with the spectral decomposition}},
	volume = {04},
	year = {2023}}

@article{Gomez:2021ujt,
	archiveprefix = {arXiv},
	author = {Gomez, Humberto and Lipinski Jusinskas, Renann and Lipstein, Arthur},
	doi = {10.1007/JHEP07(2022)004},
	eprint = {2112.12695},
	journal = {JHEP},
	pages = {004},
	primaryclass = {hep-th},
	title = {{Cosmological scattering equations at tree-level and one-loop}},
	volume = {07},
	year = {2022}}

@article{De:2023xue,
	archiveprefix = {arXiv},
	author = {De, Shounak and Pokraka, Andrzej},
	doi = {10.1007/JHEP03(2024)156},
	eprint = {2308.03753},
	journal = {JHEP},
	pages = {156},
	primaryclass = {hep-th},
	title = {{Cosmology meets cohomology}},
	volume = {03},
	year = {2024}}

@article{Arkani-Hamed:2023kig,
	archiveprefix = {arXiv},
	author = {Arkani-Hamed, Nima and Baumann, Daniel and Hillman, Aaron and Joyce, Austin and Lee, Hayden and Pimentel, Guilherme L.},
	eprint = {2312.05303},
	month = {12},
	primaryclass = {hep-th},
	title = {{Differential Equations for Cosmological Correlators}},
	year = {2023}}

@article{He:2024olr,
	archiveprefix = {arXiv},
	author = {He, Song and Jiang, Xuhang and Liu, Jiahao and Yang, Qinglin and Zhang, Yao-Qi},
	eprint = {2407.17715},
	month = {7},
	primaryclass = {hep-th},
	title = {{Differential equations and recursive solutions for cosmological amplitudes}},
	year = {2024}}

@article{Benincasa:2024ptf,
	archiveprefix = {arXiv},
	author = {Benincasa, Paolo and Brunello, Giacomo and Mandal, Manoj K. and Mastrolia, Pierpaolo and Vaz\~ao, Francisco},
	eprint = {2408.16386},
	month = {8},
	primaryclass = {hep-th},
	title = {{On one-loop corrections to the Bunch-Davies wavefunction of the universe}},
	year = {2024}}

@article{Gomez:2021qfd,
	archiveprefix = {arXiv},
	author = {Gomez, Humberto and Jusinskas, Renann Lipinski and Lipstein, Arthur},
	doi = {10.1103/PhysRevLett.127.251604},
	eprint = {2106.11903},
	journal = {Phys. Rev. Lett.},
	number = {25},
	pages = {251604},
	primaryclass = {hep-th},
	title = {{Cosmological Scattering Equations}},
	volume = {127},
	year = {2021}}

@article{Arkani-Hamed:2017fdk,
	archiveprefix = {arXiv},
	author = {Arkani-Hamed, Nima and Benincasa, Paolo and Postnikov, Alexander},
	eprint = {1709.02813},
	month = {9},
	primaryclass = {hep-th},
	title = {{Cosmological Polytopes and the Wavefunction of the Universe}},
	year = {2017}}

@article{Arkani-Hamed:2018bjr,
	archiveprefix = {arXiv},
	author = {Arkani-Hamed, Nima and Benincasa, Paolo},
	eprint = {1811.01125},
	month = {11},
	primaryclass = {hep-th},
	title = {{On the Emergence of Lorentz Invariance and Unitarity from the Scattering Facet of Cosmological Polytopes}},
	year = {2018}}

@article{Lee:2022fgr,
	archiveprefix = {arXiv},
	author = {Lee, Hayden and Wang, Xinkang},
	doi = {10.1103/PhysRevD.108.L061702},
	eprint = {2212.11282},
	journal = {Phys. Rev. D},
	number = {6},
	pages = {L061702},
	primaryclass = {hep-th},
	title = {{Cosmological double-copy relations}},
	volume = {108},
	year = {2023}}

@article{Henn:2013pwa,
	archiveprefix = {arXiv},
	author = {Henn, Johannes M.},
	doi = {10.1103/PhysRevLett.110.251601},
	eprint = {1304.1806},
	journal = {Phys. Rev. Lett.},
	pages = {251601},
	primaryclass = {hep-th},
	title = {{Multiloop integrals in dimensional regularization made simple}},
	volume = {110},
	year = {2013}}

@article{Chen:2020uyk,
	archiveprefix = {arXiv},
	author = {Chen, Jiaqi and Jiang, Xuhang and Xu, Xiaofeng and Yang, Li Lin},
	doi = {10.1016/j.physletb.2021.136085},
	eprint = {2008.03045},
	journal = {Phys. Lett. B},
	pages = {136085},
	primaryclass = {hep-th},
	title = {{Constructing canonical Feynman integrals with intersection theory}},
	volume = {814},
	year = {2021}}

@article{Chen:2022lzr,
	archiveprefix = {arXiv},
	author = {Chen, Jiaqi and Jiang, Xuhang and Ma, Chichuan and Xu, Xiaofeng and Yang, Li Lin},
	doi = {10.1007/JHEP07(2022)066},
	eprint = {2202.08127},
	journal = {JHEP},
	pages = {066},
	primaryclass = {hep-th},
	title = {{Baikov representations, intersection theory, and canonical Feynman integrals}},
	volume = {07},
	year = {2022}}

@article{Mastrolia:2018uzb,
	archiveprefix = {arXiv},
	author = {Mastrolia, Pierpaolo and Mizera, Sebastian},
	doi = {10.1007/JHEP02(2019)139},
	eprint = {1810.03818},
	journal = {JHEP},
	pages = {139},
	primaryclass = {hep-th},
	title = {{Feynman Integrals and Intersection Theory}},
	volume = {02},
	year = {2019}}

@article{Frellesvig:2019uqt,
	archiveprefix = {arXiv},
	author = {Frellesvig, Hjalte and Gasparotto, Federico and Mandal, Manoj K. and Mastrolia, Pierpaolo and Mattiazzi, Luca and Mizera, Sebastian},
	doi = {10.1103/PhysRevLett.123.201602},
	eprint = {1907.02000},
	journal = {Phys. Rev. Lett.},
	number = {20},
	pages = {201602},
	primaryclass = {hep-th},
	title = {{Vector Space of Feynman Integrals and Multivariate Intersection Numbers}},
	volume = {123},
	year = {2019}}

@article{Mizera:2019ose,
	archiveprefix = {arXiv},
	author = {Mizera, Sebastian},
	doi = {10.22323/1.383.0016},
	eprint = {2002.10476},
	journal = {PoS},
	pages = {016},
	primaryclass = {hep-th},
	title = {{Status of Intersection Theory and Feynman Integrals}},
	volume = {MA2019},
	year = {2019}}

@article{Chen:2023kgw,
	archiveprefix = {arXiv},
	author = {Chen, Jiaqi and Feng, Bo and Yang, Li Lin},
	doi = {10.1007/s11433-023-2239-8},
	eprint = {2305.01283},
	journal = {Sci. China Phys. Mech. Astron.},
	number = {2},
	pages = {221011},
	primaryclass = {hep-th},
	title = {{Intersection theory rules symbology}},
	volume = {67},
	year = {2024}}

@article{Chen:2024ovh,
    author = "Chen, Jiaqi and Feng, Bo",
    title = "{Notes on selection rules of canonical differential equations and relative cohomology}",
    eprint = "2409.12663",
    archivePrefix = "arXiv",
    primaryClass = "hep-th",
    doi = "10.1007/JHEP03(2025)009",
    journal = "JHEP",
    volume = "03",
    pages = "009",
    year = "2025"
}

@article{Chen:2023xlt,
	archiveprefix = {arXiv},
	author = {Chen, Qi and Tao, Yi-Xiao},
	doi = {10.1103/PhysRevD.108.105005},
	eprint = {2307.00870},
	journal = {Phys. Rev. D},
	number = {10},
	pages = {105005},
	primaryclass = {hep-th},
	title = {{Notes on weight-shifting operators and unifying relations for cosmological correlators}},
	volume = {108},
	year = {2023}}

@article{Tao:2022nqc,
	archiveprefix = {arXiv},
	author = {Tao, Yi-Xiao and Chen, Qi},
	doi = {10.1007/JHEP02(2023)030},
	eprint = {2210.15411},
	journal = {JHEP},
	pages = {030},
	primaryclass = {hep-th},
	title = {{A type of unifying relation in (A)dS spacetime}},
	volume = {02},
	year = {2023}}

@article{Chen:2023bji,
	archiveprefix = {arXiv},
	author = {Chen, Qi and Tao, Yi-Xiao},
	doi = {10.1007/JHEP08(2023)038},
	eprint = {2301.08043},
	journal = {JHEP},
	pages = {038},
	primaryclass = {hep-th},
	title = {{Differential operators and unifying relations for 1-loop Feynman integrands from Berends-Giele currents}},
	volume = {08},
	year = {2023}}

@article{Armstrong:2022mfr,
	archiveprefix = {arXiv},
	author = {Armstrong, Connor and Gomez, Humberto and Lipinski Jusinskas, Renann and Lipstein, Arthur and Mei, Jiajie},
	doi = {10.1103/PhysRevD.106.L121701},
	eprint = {2209.02709},
	journal = {Phys. Rev. D},
	number = {12},
	pages = {L121701},
	primaryclass = {hep-th},
	title = {{New recursion relations for tree-level correlators in anti\textendash{}de Sitter spacetime}},
	volume = {106},
	year = {2022}}

@article{Huang:2026xnq,
    author = "Huang, Li-Hong and Ma, Yan-Qing and Wang, Ziwen and Yang, Li Lin",
    title = "{Feynman integral reduction with intersection theory made simple}",
    eprint = "2604.05025",
    archivePrefix = "arXiv",
    primaryClass = "hep-th",
    month = "4",
    year = "2026"
}

@book{Arkani-Hamed:2012zlh,
    author = "Arkani-Hamed, Nima and Bourjaily, Jacob L. and Cachazo, Freddy and Goncharov, Alexander B. and Postnikov, Alexander and Trnka, Jaroslav",
    title = "{Grassmannian Geometry of Scattering Amplitudes}",
    eprint = "1212.5605",
    archivePrefix = "arXiv",
    primaryClass = "hep-th",
    reportNumber = "PUPT-2435",
    doi = "10.1017/CBO9781316091548",
    isbn = "978-1-107-08658-6, 978-1-316-57296-2",
    publisher = "Cambridge University Press",
    month = "4",
    year = "2016"
}

@article{Bern:2014kca,
    author = "Bern, Zvi and Herrmann, Enrico and Litsey, Sean and Stankowicz, James and Trnka, Jaroslav",
    title = "{Logarithmic Singularities and Maximally Supersymmetric Amplitudes}",
    eprint = "1412.8584",
    archivePrefix = "arXiv",
    primaryClass = "hep-th",
    reportNumber = "CALT-TH-2014-172, UCLA-14-TEP-109",
    doi = "10.1007/JHEP06(2015)202",
    journal = "JHEP",
    volume = "06",
    pages = "202",
    year = "2015"
}

@article{Chicherin:2018old,
    author = "Chicherin, D. and Gehrmann, T. and Henn, J. M. and Wasser, P. and Zhang, Y. and Zoia, S.",
    title = "{All Master Integrals for Three-Jet Production at Next-to-Next-to-Leading Order}",
    eprint = "1812.11160",
    archivePrefix = "arXiv",
    primaryClass = "hep-ph",
    doi = "10.1103/PhysRevLett.123.041603",
    journal = "Phys. Rev. Lett.",
    volume = "123",
    number = "4",
    pages = "041603",
    year = "2019"
}

@article{Dlapa:2021qsl,
    author = "Dlapa, Christoph and Li, Xiaodi and Zhang, Yang",
    title = "{Leading singularities in Baikov representation and Feynman integrals with uniform transcendental weight}",
    eprint = "2103.04638",
    archivePrefix = "arXiv",
    primaryClass = "hep-th",
    reportNumber = "MPP-2021-23, PCFT-21-10, USTC-ICTS-21-10",
    doi = "10.1007/JHEP07(2021)227",
    journal = "JHEP",
    volume = "07",
    pages = "227",
    year = "2021"
}

@article{Henn:2020lye,
    author = "Henn, Johannes and Mistlberger, Bernhard and Smirnov, Vladimir A. and Wasser, Pascal",
    title = "{Constructing d-log integrands and computing master integrals for three-loop four-particle scattering}",
    eprint = "2002.09492",
    archivePrefix = "arXiv",
    primaryClass = "hep-ph",
    reportNumber = "MPP-2020-13, MITP/20-006, MIT-CTP/5176",
    doi = "10.1007/JHEP04(2020)167",
    journal = "JHEP",
    volume = "04",
    pages = "167",
    year = "2020"
}

@article{Baumann:2026atn,
    author = "Baumann, Daniel and Joyce, Austin and Lee, Hayden and Salehi Vaziri, Kamran",
    title = "{Differential Equations for Massive Correlators}",
    eprint = "2604.08658",
    archivePrefix = "arXiv",
    primaryClass = "hep-th",
    month = "4",
    year = "2026"
}

@article{Hang:2024xas,
    author = "Hang, Yanfeng and Shen, Cong",
    title = "{A note on kinematic flow and differential equations for two-site one-loop graph in FRW spacetime}",
    eprint = "2410.17192",
    archivePrefix = "arXiv",
    primaryClass = "hep-th",
    doi = "10.1007/JHEP09(2025)209",
    journal = "JHEP",
    volume = "09",
    pages = "209",
    year = "2025"
}

@article{Jain:2025maa,
    author = "Jain, Diksha and Pajer, Enrico and Tong, Xi",
    title = "{Unitary and Analytic Renormalisation of Cosmological Correlators}",
    eprint = "2509.02696",
    archivePrefix = "arXiv",
    primaryClass = "hep-th",
    doi = "10.1007/JHEP03(2026)225",
    journal = "JHEP",
    volume = "03",
    pages = "225",
    year = "2026"
}

@article{Farren:2026hao,
    author = "Farren, Alexander and McCulloch, Ciaran and Pajer, Enrico and Tong, Xi",
    title = "{All-Loop Renormalization and the Phase of the de Sitter Wavefunction}",
    eprint = "2603.08794",
    archivePrefix = "arXiv",
    primaryClass = "hep-th",
    month = "3",
    year = "2026"
}

@article{Gasparotto:2024bku,
    author = "Gasparotto, Federico and Mazloumi, Pouria and Xu, Xiaofeng",
    title = "{Differential equations for tree-level cosmological correlators with massive states}",
    eprint = "2411.05632",
    archivePrefix = "arXiv",
    primaryClass = "hep-th",
    reportNumber = "MITP-24-079",
    doi = "10.1007/JHEP09(2025)043",
    journal = "JHEP",
    volume = "09",
    pages = "043",
    year = "2025"
}
\clearpage
\appendix
\begin{widetext}
\section*{Supplementary Material}
\subsection{Reduction relations and kinematic differential operators for top-sector families}
As mentioned in \cite{Chen:2023iix}, for bubble integral families involving $\theta_{12}$ or $\theta_{21}$, time IBP relations can collapse one of the time integrations and produce tadpole families. In this case, the collapsed propagator gives the following factor:
\bge
h_{n_1}^{(1)}(-k\tau_i)\,h_{n_2}^{(2)}(-k\tau_i) - h_{n_1}^{(2)}(-k\tau_i)\,h_{n_2}^{(1)}(-k\tau_i)
= (n_1-n_2) e^{ \pi \text{Im}[\nu]} \frac{  4 \ii }{\pi}  (-k\tau_i)^{-2\nu-1} \,.  \label{eq_propcontract_app}
\ede
Note that with the extra factor $e^{ \pi \text{Im}[\nu]}$, the above formula applies to both real and imaginary $\nu$.
In our integral-family language, it is implemented as a shift of the corresponding $|\mb k|$-power index by $2\nu+1$ and the $\tau$-power index by $-2\nu-1$, while the remaining families contain only one time integration. Such terms are known as remaining terms \cite{Chen:2023iix}, and will be denoted by $R$.

First, let us consider the time-IBP relations. For $\pd_{\tau_1}$, we have:
\begin{align}
	0 = & - \ii P_1 I[\{n_1,n_2,n_3,n_4\},\{a_1,a_2\},\{b_1,b_2\}] - a_1I[\{n_1,n_2,n_3,n_4\},\{a_1-1,a_2\},\{b_1,b_2\}]\nonumber \\
	    & - I[\{n_1+1,n_2,n_3,n_4\},\{a_1,a_2\},\{b_1-1,b_2\}] - I[\{n_1,n_2,n_3+1,n_4\},\{a_1,a_2\},\{b_1,b_2-1\}]\nonumber      \\
	    & -(n_1-n_2)R_1[\{n_3,n_4\},\{a_1+a_2-2\nu-1\},\{b_1+2\nu+1,b_2\}]\nonumber                \\
	    & -(n_3-n_4)R_2[\{n_1,n_2\},\{a_1+a_2-2\nu-1\},\{b_1,b_2+2\nu+1\}]\,, \label{eq:timeIBP_tau1}
\end{align}
and similarly for $\partial_{\tau_2}$:
\begin{align}
	0 = & - \ii P_2 I[\{n_1,n_2,n_3,n_4\},\{a_1,a_2\},\{b_1,b_2\}] - a_2I[\{n_1,n_2,n_3,n_4\},\{a_1,a_2-1\},\{b_1,b_2\}]\nonumber \\
	    & - I[\{n_1,n_2+1,n_3,n_4\},\{a_1,a_2\},\{b_1-1,b_2\}] - I[\{n_1,n_2,n_3,n_4+1\},\{a_1,a_2\},\{b_1,b_2-1\}]\nonumber      \\
	    & +(n_1-n_2)R_1[\{n_3,n_4\},\{a_1+a_2-2\nu-1\},\{b_1+2\nu+1,b_2\}]\nonumber                \\
	    & +(n_3-n_4)R_2[\{n_1,n_2\},\{a_1+a_2-2\nu-1\},\{b_1,b_2+2\nu+1\}]\,, \label{eq:timeIBP_tau2}
\end{align}
where $\nu$ is the common mass parameter.

As for the momentum-IBP relations, we introduce
\begin{align}
	x_1 \equiv |\mb q|\,,\qquad x_2 \equiv |\mb q+\mb k_s|\,,\qquad
	\mathcal{O}_1 \equiv \mb q^i \pd_{\mb q^i}\,,\qquad
	\mathcal{O}_2 \equiv (\mb q^i+\mb k_s^i)\pd_{\mb q^i}\,,
\end{align}
whose actions on $(x_1,x_2)$ are
\begin{align}
	\mathcal{O}_1 x_1  = x_1\,, \quad  \mathcal{O}_1 x_2  = \frac{1}{2}x_2^{-1}(x_1^2 + x_2^2 - k_s^2)\,, \nonumber\\
	 \mathcal{O}_2 x_2  = x_2 \,,   \quad \mathcal{O}_2 x_1  = \frac{1}{2}x_1^{-1}(x_1^2 + x_2^2 - k_s^2)  \,.
\end{align}
Equivalently, for any function $F(x_1,x_2)$, one has the chain rule
\begin{align}
	\mathcal{O}_\alpha F = (\mathcal{O}_\alpha x_1)\,\partial_{x_1}F + (\mathcal{O}_\alpha x_2)\,\partial_{x_2}F\,,\qquad \alpha=1,2\,.
\end{align}

From $\int \pd_{\mb q^i}[\mb q^i \cdots ]=0$ one may write an intermediate step in terms of the composite operator acting on the full integrand,
\begin{align}
	0 &= \int_{-\infty}^0 \di\tau_1\di\tau_2\int \frac{\di^d \mb q}{(2\pi)^d}\,
	\Big[d+\mathcal{O}_1\Big] \nonumber\\ 
	& \Bigg[
	e^{-\ii P_1\tau_1-\ii P_2\tau_2}
	\frac{(-\tau_1)^{a_1}(-\tau_2)^{a_2}}{x_1^{b_1}x_2^{b_2}}\,
	h_{n_1}^{(1)}(-x_1\tau_1)h_{n_2}^{(2)}(-x_1\tau_2) \times h_{n_3}^{(1)}(-x_2\tau_1)h_{n_4}^{(2)}(-x_2\tau_2)\Bigg] \nonumber\\
	& =  \int_{-\infty}^0 \di\tau_1\di\tau_2\int \frac{\di^d \mb q}{(2\pi)^d}\,
	\Bigg[d+(\mathcal{O}_1 x_1)\partial_{x_1} +(\mathcal{O}_1 x_2)\partial_{x_2}\Bigg] \nonumber\\
	&\Bigg[
	e^{-\ii P_1\tau_1-\ii P_2\tau_2}
	\frac{(-\tau_1)^{a_1}(-\tau_2)^{a_2}}{x_1^{b_1}x_2^{b_2}}\,
	h_{n_1}^{(1)}(-x_1\tau_1)h_{n_2}^{(2)}(-x_1\tau_2)
	  \times h_{n_3}^{(1)}(-x_2\tau_1)h_{n_4}^{(2)}(-x_2\tau_2)\Bigg]\,. \label{eq:momIBP_qdot_chainrule}
\end{align}
Expanding the derivatives using $\partial_x x^{-b}=-(b/x)x^{-b}$ and $\partial_x h_n(-x\tau)=(-\tau)\,h_{n+1}(-x\tau)$ yields the closed relation in terms of shifted $I$ integrals:
\begin{align}
	0 =&~ d\,I[\{n_1,n_2,n_3,n_4\},\{a_1,a_2\},\{b_1,b_2\}] - b_1\,I[\{n_1,n_2,n_3,n_4\},\{a_1,a_2\},\{b_1,b_2\}] \nonumber \\
	&- \frac12 b_2\Big(
	I[\{n_1,n_2,n_3,n_4\},\{a_1,a_2\},\{b_1,b_2\}]  + I[\{n_1,n_2,n_3,n_4\},\{a_1,a_2\},\{b_1-2,b_2+2\}] \nonumber\\
	&-k_s^2\,I[\{n_1,n_2,n_3,n_4\},\{a_1,a_2\},\{b_1,b_2+2\}]
	\Big) +I[\{n_1+1,n_2,n_3,n_4\},\{a_1+1,a_2\},\{b_1-1,b_2\}] \nonumber\\
	&+ I[\{n_1,n_2+1,n_3,n_4\},\{a_1,a_2+1\},\{b_1-1,b_2\}] +\frac12 \Big(
	I[\{n_1,n_2,n_3+1,n_4\},\{a_1+1,a_2\},\{b_1,b_2-1\}] \nonumber\\
	&+ I[\{n_1,n_2,n_3+1,n_4\},\{a_1+1,a_2\},\{b_1-2,b_2+1\}]-k_s^2\,I[\{n_1,n_2,n_3+1,n_4\},\{a_1+1,a_2\},\{b_1,b_2+1\}]
	\Big) \nonumber \\
	&+\frac12\Big(
	I[\{n_1,n_2,n_3,n_4+1\},\{a_1,a_2+1\},\{b_1,b_2-1\}]+ I[\{n_1,n_2,n_3,n_4+1\},\{a_1,a_2+1\},\{b_1-2,b_2+1\}] \nonumber\\
	&-k_s^2\,I[\{n_1,n_2,n_3,n_4+1\},\{a_1,a_2+1\},\{b_1,b_2+1\}]
	\Big)\,,
\label{eq:momIBP_qdot}
\end{align}
where $d$ is the spatial dimension (in dimensional regularization $d=3-2\ep$). The second momentum IBP can be written in the same chain-rule form as
\begin{align}
	0 ={}& \int_{-\infty}^0 \di\tau_1\di\tau_2\int \frac{\di^d \mb q}{(2\pi)^d}\,
	\Big[d+\mathcal{O}_2\Big]\nonumber\\
	&\Bigg[
	e^{-\ii P_1\tau_1-\ii P_2\tau_2}
	\frac{(-\tau_1)^{a_1}(-\tau_2)^{a_2}}{x_1^{b_1}x_2^{b_2}}\,
	h_{n_1}^{(1)}(-x_1\tau_1)h_{n_2}^{(2)}(-x_1\tau_2) \times h_{n_3}^{(1)}(-x_2\tau_1)h_{n_4}^{(2)}(-x_2\tau_2)\Bigg]\,, \label{eq:momIBP_qk_chainrule}
\end{align}
which gives the symmetric relation with the exchanges
\begin{align}
	\{n_1,n_2,n_3,n_4\} &\leftrightarrow \{n_3,n_4,n_1,n_2\}\,, \nonumber\\
	\{b_1,b_2\} &\leftrightarrow \{b_2,b_1\}\,. \nonumber
\end{align}

Note that there is another set of relations following from \eqref{eq:hdef}, called EOM relations:
\begin{align}
	I[\{2,n_2,n_3,n_4\},\{a_1,a_2\},\{b_1,b_2\}] = & - I[\{0,n_2,n_3,n_4\},\{a_1,a_2\},\{b_1,b_2\}]\nonumber      \\
   & - (2\nu+1) I[\{1,n_2,n_3,n_4\},\{a_1-1,a_2\},\{b_1+1,b_2\}]\,, \\
	I[\{n_1,2,n_3,n_4\},\{a_1,a_2\},\{b_1,b_2\}] = & - I[\{n_1,0,n_3,n_4\},\{a_1,a_2\},\{b_1,b_2\}]\nonumber      \\
& - (2\nu+1) I[\{n_1,1,n_3,n_4\},\{a_1,a_2-1\},\{b_1+1,b_2\}]\,, \\
	I[\{n_1,n_2,2,n_4\},\{a_1,a_2\},\{b_1,b_2\}] = & - I[\{n_1,n_2,0,n_4\},\{a_1,a_2\},\{b_1,b_2\}]\nonumber      \\
     & - (2\nu+1) I[\{n_1,n_2,1,n_4\},\{a_1-1,a_2\},\{b_1,b_2+1\}]\,, \\
	I[\{n_1,n_2,n_3,2\},\{a_1,a_2\},\{b_1,b_2\}] = & - I[\{n_1,n_2,n_3,0\},\{a_1,a_2\},\{b_1,b_2\}]\nonumber      \\
     & - (2\nu+1) I[\{n_1,n_2,n_3,1\},\{a_1,a_2-1\},\{b_1,b_2+1\}]\,,
\end{align}
where $\nu$ is the common mass parameter.

In addition to the above IBP relations, we also have some symmetry relations:
\begin{align}
	I[\{n_1,n_2,n_3,n_4\},\{a_1,a_2\},\{b_1,b_2\}] = I[\{n_3,n_4,n_1,n_2\},\{a_1,a_2\},\{b_2,b_1\}]\,.
\end{align}

As our final step, we will present the differential equations of this integral family. Although we set $P_1=P_2$, here we give the $P_1$ and $P_2$ derivatives separately:
\begin{align}
	\frac{\partial}{\partial P_1}I[\{n_1,n_2,n_3,n_4\},\{a_1,a_2\},\{b_1,b_2\}] &= \ii\,I[\{n_1,n_2,n_3,n_4\},\{a_1+1,a_2\},\{b_1,b_2\}]\,, \label{eq:GdP1}\\
	\frac{\partial}{\partial P_2}I[\{n_1,n_2,n_3,n_4\},\{a_1,a_2\},\{b_1,b_2\}] &= \ii\,I[\{n_1,n_2,n_3,n_4\},\{a_1,a_2+1\},\{b_1,b_2\}]\,. \label{eq:GdP2}
\end{align}
From
\begin{align}
	\frac{\partial}{\partial k_s}|\mb q+\mb k_s|
	=\frac{k_s^2+\mb k_s\cdot \mb q}{k_s|\mb q+\mb k_s|}
	=\frac{k_s^2+|\mb q+\mb k_s|^2-|\mb q|^2}{2k_s|\mb q+\mb k_s|}\,,
\end{align}
the $k_s$-derivative gives:
\ie
	&\frac{\partial}{\partial k_s}I[\{n_1,n_2,n_3,n_4\},\{a_1,a_2\},\{b_1,b_2\}] \\
	=&-\frac{b_2}{2k_s}\Big(
	k_s^2\,I[\{n_1,n_2,n_3,n_4\},\{a_1,a_2\},\{b_1,b_2+2\}]
	+I[\{n_1,n_2,n_3,n_4\},\{a_1,a_2\},\{b_1,b_2\}] \\
	&
	-I[\{n_1,n_2,n_3,n_4\},\{a_1,a_2\},\{b_1-2,b_2+2\}]
	\Big)+\frac{1}{2k_s}\Big(
	k_s^2\,I[\{n_1,n_2,n_3+1,n_4\},\{a_1+1,a_2\},\{b_1,b_2+1\}] \\
	&
	+I[\{n_1,n_2,n_3+1,n_4\},\{a_1+1,a_2\},\{b_1,b_2-1\}]
	-I[\{n_1,n_2,n_3+1,n_4\},\{a_1+1,a_2\},\{b_1-2,b_2+1\}]
	\Big) \\
	&+\frac{1}{2k_s}\Big(
	k_s^2\,I[\{n_1,n_2,n_3,n_4+1\},\{a_1,a_2+1\},\{b_1,b_2+1\}]
	+I[\{n_1,n_2,n_3,n_4+1\},\{a_1,a_2+1\},\{b_1,b_2-1\}] \\
	&
	-I[\{n_1,n_2,n_3,n_4+1\},\{a_1,a_2+1\},\{b_1-2,b_2+1\}]
	\Big)\,. \label{eq:Gdks}
\fe

The family also satisfies the scaling relation:
\begin{align}
	\left( k_s \frac{\partial}{\partial k_s} + P_1 \frac{\partial}{\partial P_1} + P_2 \frac{\partial}{\partial P_2} \right) I[\{\bm n\},\{a_1,a_2\},\{b_1,b_2\}] \n\\
	= (d - b_1 - b_2 - a_1 - a_2 - 2) I[\{\bm n\},\{a_1,a_2\},\{b_1,b_2\}]\,. \label{eq:Gscaling}
\end{align}

In conclusion, the IBP relations for the top sector are \eqref{eq:timeIBP_tau1}, \eqref{eq:timeIBP_tau2}, \eqref{eq:momIBP_qdot}, and \eqref{eq:momIBP_qk_chainrule} together with the symmetry relations and the EOM relations. The differential equations for the top sector are \eqref{eq:GdP1}, \eqref{eq:GdP2}, and \eqref{eq:Gdks}. The scaling relation \eqref{eq:Gscaling} is used as a consistency check on the differential equations.

\subsection{Reduction relations and kinematic differential operators for tadpole families $R_1,R_2$}
In the IBP relations of the bubble diagram, tadpole integral families appear as remaining terms. 
\begin{align}
R_1[\{n_3,n_4\},\{a\},\{b_1,b_2\}]
\equiv {}&
e^{\pi\text{Im}[\nu]}\frac{4\ii}{\pi}
\int_{-\infty}^0\di\tau
\int\frac{\di^d\mb q}{(2\pi)^d}\,
e^{-\ii P_0\tau}
\frac{(-\tau)^a}
{|\mb q|^{b_1}|\mb q+\mb k_s|^{b_2}}
\nonumber\\
&\times\frac12\sum_{r=0,1}
h_{n_3}^{(1+r)}(-|\mb q+\mb k_s|\tau)
h_{n_4}^{(2-r)}(-|\mb q+\mb k_s|\tau)\,,
\\
R_2[\{n_1,n_2\},\{a\},\{b_1,b_2\}]
\equiv {}&
e^{\pi\text{Im}[\nu]}\frac{4\ii}{\pi}
\int_{-\infty}^0\di\tau
\int\frac{\di^d\mb q}{(2\pi)^d}\,
e^{-\ii P_0\tau}
\frac{(-\tau)^a}
{|\mb q|^{b_1}|\mb q+\mb k_s|^{b_2}}
\nonumber\\
&\times\frac12\sum_{r=0,1}
h_{n_1}^{(1+r)}(-|\mb q|\tau)
h_{n_2}^{(2-r)}(-|\mb q|\tau)\,.
\end{align}

Such terms have their own IBP relations. Here we only give the IBP relations for $R_1$, and the corresponding relations for $R_2$ follow via the symmetry $\mb q \to -(\mb q+\mb k_s)$, which exchanges $b_1\leftrightarrow b_2$ and relabels Hankel indices appropriately.
\begin{align}
	R_2[\{n_1,n_2\},\{a\},\{b_1,b_2\}] = R_1[\{n_1,n_2\},\{a\},\{b_2,b_1\}]. \label{eq:R1R2sym}
\end{align}
The family \(R\) defined in the main text is \(R_2\). Equivalently,
using \eqref{eq:R1R2sym},
\[
R[\{\bm n\},\{a\},\{b_1,b_2\}]
=
R_2[\{\bm n\},\{a\},\{b_1,b_2\}]
=
R_1[\{\bm n\},\{a\},\{b_2,b_1\}] .
\]
We therefore give the reduction relations only for \(R_1\).
The time IBP relation for $R_1$ is
\begin{align}
	0 = & -\ii(P_1+P_2) R_1[\{n_3,n_4\},\{a\},\{b_1,b_2\}] - a R_1[\{n_3,n_4\},\{a-1\},\{b_1,b_2\}] \nonumber       \\
	    & - R_1[\{n_3+1,n_4\},\{a\},\{b_1,b_2-1\}] - R_1[\{n_3,n_4+1\},\{a\},\{b_1,b_2-1\}]\,, \label{eq:R1timeIBP}
\end{align}
and two momentum IBP relations read
\begin{align}
	0 = &~ d R_1[\{n_3,n_4\},\{a\},\{b_1,b_2\}] - b_1 R_1[\{n_3,n_4\},\{a\},\{b_1,b_2\}] \nonumber                    \\
	    & - \frac12 b_2\Big( R_1[\{n_3,n_4\},\{a\},\{b_1,b_2\}] + R_1[\{n_3,n_4\},\{a\},\{b_1-2,b_2+2\}] -k_s^2 R_1[\{n_3,n_4\},\{a\},\{b_1,b_2+2\}] \Big) \nonumber                                                \\
	    & + \frac12 \Big(
	    R_1[\{n_3+1,n_4\},\{a+1\},\{b_1,b_2-1\}] + R_1[\{n_3+1,n_4\},\{a+1\},\{b_1-2,b_2+1\}] \nonumber\\
	    &-k_s^2 R_1[\{n_3+1,n_4\},\{a+1\},\{b_1,b_2+1\}]
	    \Big)+ \frac12 \Big(
	    R_1[\{n_3,n_4+1\},\{a+1\},\{b_1,b_2-1\}] \nonumber\\
     &+ R_1[\{n_3,n_4+1\},\{a+1\},\{b_1-2,b_2+1\}]-k_s^2 R_1[\{n_3,n_4+1\},\{a+1\},\{b_1,b_2+1\}]
	    \Big)\,. \label{eq:R1momIBP}
    \end{align}
\begin{align}
	0 = & d R_1[\{n_3,n_4\},\{a\},\{b_1,b_2\}] - b_2 R_1[\{n_3,n_4\},\{a\},\{b_1,b_2\}] \nonumber                    \\
	    & - \frac12 b_1\Big( R_1[\{n_3,n_4\},\{a\},\{b_1,b_2\}] + R_1[\{n_3,n_4\},\{a\},\{b_1+2,b_2-2\}] \nonumber     \\
	    & -k_s^2 R_1[\{n_3,n_4\},\{a\},\{b_1+2,b_2\}] \Big)+ R_1[\{n_3+1,n_4\},\{a+1\},\{b_1,b_2-1\}] \nonumber\\
     &+ R_1[\{n_3,n_4+1\},\{a+1\},\{b_1,b_2-1\}]\,. \label{eq:R1momIBP2}
\end{align}
These two relations come from $\int \pd_{\mb q^i}[\mb q^i\cdots]=0$ and $\int \pd_{\mb q^i}[(\mb q^i+\mb k_s^i)\cdots]=0$, respectively.

Note that in the tadpole case, we also have relations coming from \eqref{eq:hdef}:
\begin{align}
	R_1[\{2,n_4\},\{a\},\{b_1,b_2\}] = & - R_1[\{0,n_4\},\{a\},\{b_1,b_2\}] \nonumber                            \\
 & - (2\nu+1) R_1[\{1,n_4\},\{a-1\},\{b_1,b_2+1\}]\,, \label{eq:R1eom1} \\
	R_1[\{n_3,2\},\{a\},\{b_1,b_2\}] = & - R_1[\{n_3,0\},\{a\},\{b_1,b_2\}] \nonumber                            \\
& - (2\nu+1) R_1[\{n_3,1\},\{a-1\},\{b_1,b_2+1\}]\,. \label{eq:R1eom2}
\end{align}
and the following symmetry relation
\begin{align}
	 R_1[\{n_3,n_4\},\{a\},\{b_1,b_2\}] = R_1[\{n_4,n_3\},\{a\},\{b_1,b_2\}] \label{eq:R1sym}\,. 
\end{align}

Let us turn to the differential equations. Taking $P_1=P_2=P_0/2$, the $P_0$-derivative gives:
\begin{align}
	\frac{\partial}{\partial P_0} R_1[\{n_3,n_4\},\{a\},\{b_1,b_2\}] = \ii R_1[\{n_3,n_4\},\{a+1\},\{b_1,b_2\}]\,. \label{eq:R1dP0}
\end{align}
and the $k_s$-derivative gives
\begin{align}
	&\frac{\partial}{\partial k_s} R_1[\{n_3,n_4\},\{a\},\{b_1,b_2\}] \n\\
	=&-\frac{b_2}{2k_s}\Big(
	k_s^2\,R_1[\{n_3,n_4\},\{a\},\{b_1,b_2+2\}] 
	+R_1[\{n_3,n_4\},\{a\},\{b_1,b_2\}] 
	-R_1[\{n_3,n_4\},\{a\},\{b_1-2,b_2+2\}]
	\Big) \nonumber\\
	&+\frac{1}{2k_s}\Big(
	k_s^2\,R_1[\{n_3+1,n_4\},\{a+1\},\{b_1,b_2+1\}] 
	+R_1[\{n_3+1,n_4\},\{a+1\},\{b_1,b_2-1\}] \nonumber\\
	&
	-R_1[\{n_3+1,n_4\},\{a+1\},\{b_1-2,b_2+1\}]
	\Big)+\frac{1}{2k_s}\Big(
	k_s^2\,R_1[\{n_3,n_4+1\},\{a+1\},\{b_1,b_2+1\}] 
	\nonumber\\
 &+R_1[\{n_3,n_4+1\},\{a+1\},\{b_1,b_2-1\}] 
	-R_1[\{n_3,n_4+1\},\{a+1\},\{b_1-2,b_2+1\}]
	\Big)\,. \label{eq:R1dks}
\end{align}

Similarly to the bubble case, we have the following scaling relation:
\begin{align}
	\left( k_s \frac{\partial}{\partial k_s} + P_0 \frac{\partial}{\partial P_0} \right) R_1[\{n_3,n_4\},\{a\},\{b_1,b_2\}] = (d - b_1 - b_2 - a - 1) R_1[\{n_3,n_4\},\{a\},\{b_1,b_2\}]\,, \label{eq:R1scaling}
\end{align}
which will be used to verify the correctness of the differential equations.

In conclusion, the IBP relations for the remaining terms are \eqref{eq:R1timeIBP}, \eqref{eq:R1momIBP}, and \eqref{eq:R1momIBP2} together with symmetry relations and EOM relations. The differential equations are \eqref{eq:R1dP0} and \eqref{eq:R1dks}. As before, the remaining terms also satisfy the scaling relation \eqref{eq:R1scaling}.

\subsection{Details for $\di \log$-form building blocks involving $\Omega^\tau$}
The first expression below gives the logarithmic one-form used
in class (b), while the remaining identities justify those used
in classes (d) and (e). Since $\Omega_{ex}$ contains only
$\log(P_i\pm x_1\pm x_2)$, the required combinations can be written as
\begin{align}
\di \big[\int(\partial_{P_2}\Omega_{ex})_{\bm{m}\bm{n}}\di x_2\big]&=\bigg[\big[(-1)^{\lfloor |\bm{m}|\rfloor/2}\frac{P_2^2-x_1^2-x_2^2}{2x_1x_2}\partial_{P_2}\Omega_{ex}\big]_{\bm{m}\bm{n}}+(\Omega_{re})_{\bm{m}\bm{n}}\bigg]\,\di x_1+(\partial_{P_2}\Omega_{ex})_{\bm{m}\bm{n}}\,\di x_2,
\nonumber\\
\partial_{x_j}\Omega_{ex}\,\di x_i
&=
\left(
\partial_{x_j}\Omega^\tau
-\partial_{x_j}\Omega_{x_j}
\right)\di x_i\,.
\label{dOmega}
\end{align}
where $\Omega_{re}$ is given in the supplementary files. Here and below, $\hat I$ denotes the corresponding Baikov form before the $x_1$ and $x_2$ integrations. Using
\begin{align}
& \hat{I}[\{\bm{n}\},\{a_1+1,a_2\},\{b_1,b_2\}]= -\ii \partial_{P_1}\hat{I}[\{\bm{n}\},\{a_1,a_2\},\{b_1,b_2\}]
=-\ii \left(\partial_{P_1}\Omega^\tau\right)_{\bm n\bm m} \hat{I}[\{\bm{m}\},\{a_1,a_2\},\{b_1,b_2\}]\n\\
& \quad\quad\quad\quad\quad\quad\quad\quad\quad\quad\quad\quad =-\ii \left(\partial_{P_1}\Omega_{ex}\right)_{\bm n\bm m} \hat{I}[\{\bm{m}\},\{a_1,a_2\},\{b_1,b_2\}] \n \\
& (b_0+1)\hat{I}[\{\bm{n}\},\{0,0\},\{1,3\}] =\left(\partial_{x_2}\Omega^\tau\right)_{\bm n\bm m} \hat{I}[\{\bm{m}\},\{0,0\},\{1,2\}] - \ep \partial_{x_2} \log \mG  \hat{I}[\{\bm{n}\},\{0,0\},\{1,2\}],\n\\
& ~~~~~~~~~~~~~~  =\left( \partial_{x_2}\Omega_{ex} + \partial_{x_2}\Omega_{x_2} \right)_{\bm n\bm m} \hat{I}[\{\bm{m}\},\{0,0\},\{1,2\}] - \ep \frac{ \partial_{x_2} \mG }{\mG x_2} \hat{I}[\{\bm{n}\},\{0,0\},\{1,1\}]
\end{align}
and notice that the $\partial \mG$ term is also a $\di \log$-form
\begin{align*}
\frac{ \partial_{x_2}\mG \di x_1 \di x_2}{\mG x_2}=2\left[\di \tanh ^{-1}\left(\frac{x_1}{k_s+x_2}\right)- \di \tanh
   ^{-1}\left(\frac{x_1}{k_s-x_2}\right)\right]\di \log x_2 \,.
\end{align*}
One finds $\partial_{P_j}\Omega_{ex}\,\di x_i\propto\tau_j\,\di x_i$ and
$\partial_{x_j}\Omega_{ex}\,\di x_i ~ \di \log x_j + (\text{$\partial \mG$ term}) \propto \frac{1}{x_j^2}\di x_i \di x_j$
as additional $\di\log$-form building blocks. Here \(\propto\) is understood up to IBP equivalence.

\subsection{Raising $d=1$ back to $d=3$}
In this section, we demonstrate that all dimension-1 master integrals can be raised to their corresponding dimension-3 master integrals. Therefore, they do not require special treatment in the differential equations. 

For the normalized Baikov measure, one has
\begin{align}
	\frac{2}{\pi^{d/2}}\,\di^d \mb q = C_d\,\mG^{\frac{d-3}{2}}\mK^{-\frac{d-2}{2}}\,\di z_1\,\di z_2\,.
\end{align}
In particular, for $d=1-2\ep$ and $d=3-2\ep$,
\begin{align}\label{shift}
	C_{1-2\ep}\,\mG^{-\ep-1}\mK^{\ep+1/2}
	= \partial_{q^2}\!\left[C_{3-2\ep}\,\mK^{-1/2+\ep}\mG^{-\ep}\right]\,,
\end{align}
where $\partial_{q^2}\equiv\partial_{z_1}+\partial_{z_2}$ and we used $\partial_{q^2}\mG=\mK$ together with $\Gamma(1-\ep)=-\ep\,\Gamma(-\ep)$.

Therefore, for
\begin{align}
	\hat{I}[\{\bm n\},\{a_1,a_2\},\{b_1,b_2\};d=1]
	=&~\di z_1\,\di z_2\,z_1^{-(b_0+b_1)/2}z_2^{-(b_0+b_2)/2}\,I^{\tau}_{\bm n}\,\n\\
	&\times (-\tau_1)^{a_1}(-\tau_2)^{a_2}\,C_{1-2\ep}\,\mG^{-\ep-1}\mK^{\ep+1/2}\,,
\end{align}
Integration by parts gives
\begin{align}
	\hat{I}[\{\bm n\},\{a_1,a_2\},\{b_1,b_2\};d=1]
	=&~\di z_1\,\di z_2\,\Bigg[\left(\frac{b_0+b_1}{2z_1}+\frac{b_0+b_2}{2z_2}\right)I^{\tau}_{\bm n}
	-\left(\partial_{z_1}+\partial_{z_2}\right)I^{\tau}_{\bm n}\Bigg] \n\\
	&\times (-\tau_1)^{a_1}(-\tau_2)^{a_2}\,z_1^{-(b_0+b_1)/2}z_2^{-(b_0+b_2)/2}\,
	C_{3-2\ep}\,\mG^{-\ep}\mK^{-1/2+\ep}\,.
\end{align}
Equivalently, in $(x_1,x_2)$ variables with $\partial_{q^2}=\frac{1}{2x_1}\partial_{x_1}+\frac{1}{2x_2}\partial_{x_2}$,
\begin{align}
	\hat{I}[\{\bm n\},\{a_1,a_2\},\{b_1,b_2\};d=1]
	=&~2\,\di x_1\,\di x_2\,\Bigg[\left(\frac{b_0+b_1}{x_1^2}+\frac{b_0+b_2}{x_2^2}\right)I^{\tau}_{\bm n}
	-\left(\frac{1}{x_1}\partial_{x_1}+\frac{1}{x_2}\partial_{x_2}\right)I^{\tau}_{\bm n}\Bigg] \n\\
	&\times (-\tau_1)^{a_1}(-\tau_2)^{a_2}\,x_1^{-(b_0+b_1-1)}x_2^{-(b_0+b_2-1)} \,
	C_{3-2\ep}\,\mG^{-\ep}\mK^{-1/2+\ep}\,.
\end{align}
In particular, for the $d=1$ candidates above,
\begin{align}
	\hat{I}[\{\bm n\},\{0,0\},\{1,0\};d=1]
	&= \di z_1\,\di z_2\,z_1^{-b_0/2-1/2}z_2^{-b_0/2}\,I^{\tau}_{\bm n}\,\mG^{-\ep-1}\mK^{\ep+1/2} \n\\
	&=2\,\di x_1\,\di x_2\,\Bigg[\left(\frac{b_{0}+1}{x_1^2}+\frac{b_{0}}{x_2^2}\right)I^{\tau}_{\bm n}
	-\left(\frac{1}{x_1}\partial_{x_1}+\frac{1}{x_2}\partial_{x_2}\right)I^{\tau}_{\bm n}\Bigg]\n\\
	&\times x_1^{-b_0}x_2^{-b_0+1}\,\mG^{-\ep}\mK^{-1/2+\ep}, \n\\
	\hat{I}[\{\bm n\},\{1,0\},\{0,0\};d=1]
	&=2\,\di x_1\,\di x_2\,\Bigg[b_0\left(\frac{1}{x_1^2}+\frac{1}{x_2^2}\right)I^{\tau}_{\bm n}
	-\left(\frac{1}{x_1}\partial_{x_1}+\frac{1}{x_2}\partial_{x_2}\right)I^{\tau}_{\bm n}\Bigg]\n\\
	&\times (-\tau_1)\,x_1^{-b_0+1}x_2^{-b_0+1}\,\mG^{-\ep}\mK^{-1/2+\ep}\,.
\end{align}
Hence, all $d=1$ master integrals can be rewritten as $d=3$ master integrals.

\end{widetext}

\end{document}